\documentclass[
reprint,
amsmath,assume,
aps
]{revtex4-2}
\usepackage[dvipsnames]{xcolor}
\usepackage{graphicx}
\usepackage{subcaption}
\usepackage{derivative}
\usepackage{dcolumn}
\usepackage{wrapfig}
\usepackage{bm}
\usepackage{hyperref}

\usepackage{xcolor}
\usepackage{xparse}
\usepackage{amsfonts}

\ExplSyntaxOn

\keys_define:nn { miguel/label }
 {
  label   .tl_set:N = \l_miguel_label_tl,
  unknown .code:n   = \clist_put_right:Nx \l_miguel_label_clist
                       { \l_keys_key_tl = \exp_not:n { #1 } }
 }
\clist_new:N \l_miguel_label_clist
\box_new:N \l_miguel_label_box
\box_new:N \l_miguel_label_image_box

\NewDocumentCommand{\xincludegraphics}{O{}m}
 {
  \tl_clear:N \l_miguel_label_tl
  \clist_clear:N \l_miguel_label_clist
  \keys_set:nn { miguel/label } { #1 }
  \tl_if_empty:NTF \l_miguel_label_tl
   {
    \miguel_includegraphics:Vn \l_miguel_label_clist { #2 }
   }
   {
    \hbox_set:Nn \l_miguel_label_image_box
     {
      \miguel_includegraphics:Vn \l_miguel_label_clist { #2 }
     }
    \hbox_set:Nn \l_miguel_label_box
     {
      \skip_horizontal:n { -3pt }
      \fcolorbox{white}{white}{\footnotesize \tl_use:N \l_miguel_label_tl}
     }
    \leavevmode
    \box_use:N \l_miguel_label_image_box
    \skip_horizontal:n { -\box_wd:N \l_miguel_label_image_box }
    \hbox_overlap_right:n
     {
      \box_move_up:nn
       {
        \box_ht:N \l_miguel_label_image_box - 
        \box_ht:N \l_miguel_label_box - -3pt
       }
       { \box_use_drop:N \l_miguel_label_box }
     }
    \skip_horizontal:n { \box_wd:N \l_miguel_label_image_box }
   }
 }

\cs_new_protected:Nn \miguel_includegraphics:nn
 {
  \includegraphics[#1]{#2}
 }
\cs_generate_variant:Nn \miguel_includegraphics:nn { V }

\ExplSyntaxOff

\begin{document}

\preprint{APS/123-QED}

\title{Phase transitions in $q$-state clock model}

\author{Arpita Goswami}
\author{Ravi Kumar}
\author{Monikana Gope}
\author{Shaon Sahoo}%
 \email{Corresponding author: shaon@iittp.ac.in}
\affiliation{%
 Indian Institute of Technology Tirupati, India, 517619 \\
}%


\begin{abstract}
The $q-$state clock model, sometimes called the discrete $XY$ model, is known to show a second-order (symmetry breaking) phase transition in two-dimension (2D) for $q\le 4$ ($q=2$ corresponds to the Ising model). On the other hand, the $q\to\infty$ limit of the model corresponds to the $XY$ model, which shows the infinite order (non-symmetry breaking) Berezinskii-Kosterlitz-Thouless (BKT) phase transition in 2D. Interestingly, the 2D clock model with $q\ge 5$ is predicted to show three different phases and two associated phase transitions. There are varying opinions about the actual characters of phases and the associated transitions. In this work, we develop the basic and higher-order mean-field (MF) theories to study the $q$-state clock model systematically. Our MF calculations reaffirm that, for large $q$, there are three phases: (broken) $\mathbb{Z}_q$ symmetric ferromagnetic phase at the low temperature, emergent $U(1)$ symmetric BKT phase at the intermediate temperature, and paramagnetic (disordered) phase at the high temperature. The phase transition at the higher temperature is found to be of the BKT type, and the other transition at the lower temperature is argued to be a large-order spontaneous symmetry-breaking (SSB) type (the largeness of transition order yields the possibility of having some of the numerical characteristics of a BKT transition). The higher-order MF theory developed here better characterizes phases by estimating the spin-spin correlation between two neighbors.    
\end{abstract}


 \maketitle
\section{Introduction} \label{sec1}
The two-dimensional (2D) Ising model shows a spontaneous symmetry-breaking (SSB) phase transition at a finite temperature \cite{PhysRev.65.117,beale2021statistical}. The 2D $XY$ model does not show this SSB transition due to the restriction imposed by the Mermin-Wagner theorem \cite{mermin}. This model, however, shows a different type of phase transition, known as Berezinskii-Kosterlitz-Thouless or BKT transition \cite{berezinskii1971destruction,kosterlitz2016kosterlitz, kosterlitz,KL,gupta1988phase}, due to the unbinding of bound vortexes-antivortexes pairs (topological defects). The $q-$state clock model, a generalization of the Potts model \cite{baxter1973potts,RevModPhys.54.235} and one of the different possible discretizations of the $XY$ model \cite{vanEnter11,Butt23}, represents the Ising and the $XY$ models in two different limits. The clock model becomes the $\mathbb{Z}_2$ symmetric Ising model when $q=2$ and it represents the $U(1)$ symmetric $XY$ model in the $q\to \infty$ limit  \cite{chen2017phase, negrete2021short, Li2020}. The clock model exhibits rich physics in terms of the nature of its phases and the associated critical phenomena. 

Many questions arise while studying the 2D clock model. For example, since the Ising and the $XY$ model show two different types of phase transitions, it is of interest to know what happens to the 2D clock model when $q$ is finite and $q>2$. It is generally agreed that there is some critical value $q_c$; the 2D clock model shows only one phase transition (of the SSB type) when $q<q_c$. On the other hand, when $q\ge q_c$, the model is predicted to show two phase transitions \cite{chen2017phase, negrete2021short, Li2020,pre106.024106}. It is now generally accepted that $q_c=5$ \cite{Elitzur79,Tobochnik82}. 

The next issue is about the nature of phases and the associated phase transitions. For a finite value of $q$, the Mermin-Wagner theorem \cite{mermin} does not apply to the clock model. In addition, the non-symmetry breaking BKT phase transition is originally shown to exist for the $XY$ model where spins have continuous rotational symmetry $U(1)$. This raises the question on whether the phase transitions observed in the 2D clock model are of SSB type, especially for small values of $q$ (say, 5 or 6). In this context, it is then somewhat counterintuitive that a large number of the recent studies point out that the $\mathbb{Z}_q$ symmetric clock model can exhibit the BKT type transition \cite{Li2020,Hong2020,ortiz2012,Surungan2019,Kumano13}. The counterclaims about the nature of the transition(s) do also exist, especially for the small values of $q$ \cite{Lapilli06,Baek13,Hwang09}. We note that the location and the nature of transition(s) can be identified in different ways: the finite size scaling behavior of quantities like the helicity modulus \cite{Kumano13,Lapilli06} and Fisher zero \cite{Hong2020,Hwang09}, and also from the emergence of U(1) symmetry (identified by, for example, the scale invariance of the correlation ratio \cite{Surungan2019}). The results may vary based on the different quantities studied, besides the accuracy of the method used. In this paper, we find that the transition at the higher temperature is of the BKT type, in agreement with the current consensus, and we argue that the transition at the lower temperature is a large-order spontaneous symmetry-breaking (SSB) transition.

Although there are varying opinions on the nature of the phase transitions, it is generally accepted that there are three phases for $q\ge 5$ \cite{Li2020,ortiz2012,Surungan2019,Alcaraz80}, which include an ordered ferromagnetic phase at low temperature and a disordered paramagnetic phase at high temperature. There is an intermediate phase which is predicted to be a quasiliquid (sometimes called the BKT phase). It is important to investigate how the $U(1)$ symmetry arises from a $\mathbb{Z}_q$ symmetric spin model and how the BKT phase forms (in terms of the vortexes-anti vortexes pairs) with this emergent symmetry. Since many of the issues in this field are not yet completely resolved and some results require more clarity, it is worthwhile to re-investigate and gain more insight into the nature of the phases and associated transitions. To this effort, we develop a mean-field (MF) theory (both basic and higher order) for the $q-$state clock model. To the best of our knowledge, ours is the first systematic mean-field study of the $q$-state clock model. 

The MF theory is an important approach to analyze properties of model systems \cite{beale2021statistical}. Its success in analyzing classical or quantum models is well known. The MF theories work better in the higher dimensions and/ or when the coordination number is large. Although, in this work, we are mainly interested in the two-dimensional systems, we expect to get reasonable results in the large $q$ limit. This is because the change in the mean field will be much less due to elementary excitation or low-temperature fluctuation in the neighbors. Additionally, since the MF theory works better in higher dimensions, our MF theory can help us analyze properties of higher dimensional (say, 3-dimensional) $q$-state clock model.    

Our main contributions in this paper include the following. We develop the basic and higher order MF theories for the $q$-state clock model and our analysis reaffirms the existence of three separate phases in 2D clock model (for large $q$). We demonstrate how the $U(1)$ symmetry arises in the intermediate temperature regime for a Hamiltonian with $\mathbb{Z}_q$ symmetry. We find that the transition at the higher temperature is of BKT type, and the transition temperature is almost $q$-independent, i.e., $T_{BKT} = O(1)$. We argue that the transition at the lower temperature is a large-order SSB transition with possibility of having some numerical characteristics of a BKT transition. This transition temperature decreases as $1/q^2$, i.e., $T_{SSB}=O(q^{-2})$.  

This article is arranged in the following way. In Sec. \ref{sec2}, we develop the basic (zeroth order) MF theory and analyze the 2D clock model using the theory. Next, in Sec. \ref{sec3}, we construct the first-order MF theory which gives us better insight into the phases of the model. Using this higher-order theory, we estimate the nearest-neighbor spin-spin correlation and find a slightly better estimation of the BKT transition temperature. We conclude our work in Sec. \ref{sec4}. 

\section{Mean-Field Theory} \label{sec2}
 The $q-$state clock model represents a spin system where the spins lie in the $XY$ plane and can orient only in $q$ possible directions. This model, without a magnetic field, is represented by the following Hamiltonian:
\begin{equation}
H=-J\sum_{\langle ij \rangle }{\cos{(\theta_i-\theta_j)}},
\label{H_clc}
\end{equation}
where $J$ is the nearest-neighbor exchange constant (taken to be positive) and $\langle{ij}\rangle$ represents summation over the nearest neighbors. Here $\theta_i$ is the angle between the $i$-th spin and the reference, say $X$, axis. This angle takes $q$ possible values $\theta_i=2\pi k_i/q$ where $k_i$ can take any integer value from 1 to $q$. As the exact results for this model are scarce, it is important to develop an appropriate mean-field approach to understand the behavior of this model. 
 
 As a first step in developing the (basic or zeroth order) mean-field theory for this model, we define a {\it complex spin variable} $S_k$ corresponding to the $k$-th spin: $S_k = e^{i\theta_k}$, where $\theta_k=2\pi j_k/q$. Here $j_k$ takes any integer value from 1 to $q$. With this complex spin variable, we can rewrite the Hamiltonian in  Eq. \ref{H_clc} as,    
 \begin{equation}
    H=-J Re{\sum{S_i S^*_j}}.
\label{H_cmx}
\end{equation}
We now perform the basic (zero-th order) MF theory over the complex spin variables. In the MF approach, each variable ($S_i$) is replaced by its mean value plus a small fluctuation term ($\delta S_i$), 
\[
S_i=\langle S_i \rangle +\delta S_i.
\]
If we define the complex mean-field parameter as $l=\sum_{i=1}^{N}{\frac{\langle S_i \rangle}{N}}$, for translationally invariant system, we can write  
 \[
S_i S^*_j=(l+\delta S_i)(l+\delta S_j)^*\approx |l|^2+l\delta S_j^* + l^*\delta S_i,
\]
after neglecting the small (second order) term $\delta S_i \delta S^*_j$. After replacing $\delta S_i$ by $S_i-l$, we get the following mean-field Hamiltonian from Eq. \ref{H_cmx},
\begin{equation} \nonumber
H_{MF}=-J Re{\sum_{\langle ij \rangle}
{(|l|^2+l^* (S_i-l)+ l (S_j-l)^*)}}.
\end{equation}
We here note that $\sum_{\langle ij \rangle }S_i=\sum_{\langle ij \rangle }S_j$ and $\sum_{\langle ij \rangle}=
\frac{1}{2}\sum_{i=1}^{N} \sum_{j\in(nn)}$, where $\sum_{j\in(nn)}$ represents the sum over the nearest neighbors, $N$ is the total number of lattice sites. With these, we get 
\begin{equation}
H_{MF}=-J\frac{1}{2}Re(-
{{Nr{|l|}^2+r\sum_{i=1}^{N}(l^*S_i+l S^*_i))}},
\label{H_mf1}
\end{equation}
where $r$ is the number of nearest neighbors. As  $l^*S_i+l S^*_i=2Re(l^* S_i)$, we can write
\begin{equation}
H_{MF}=\frac{1}{2}
{{NJr {|l|}^2-Jr\sum_{i=1}^{N}Re(l^*S_i)}}.
\label{H_mf2}
\end{equation}
The mean-field parameter $l$ is a complex quantity, which we now write in terms of two real parameters: $l=l_1+i l_2$. Physically, $l_1$ and $l_2$ represent the average values of the $X$ and the $Y$ components of a spin. With $S_j=\cos{\frac{2 \pi j_k}{q}}+i \sin{\frac{2 \pi j_k}{q}}$, we can recast Eq. \ref{H_mf2} as
\begin{equation}
H_{MF}=\frac{1}{2} NJr(l_1^2+l_2^2)- Jr\sum_{i=1}^{N}(l_1 \cos{\frac{2 \pi i_k}{q}}+l_2\sin{\frac{2 \pi i_k}{q}}).
\label{H_mf3}
\end{equation}
It may be mentioned again here that $i_k$ can take any of the $q$ values from 1 to $q$. If we write $H_{MF}=\sum_i^N h_i$, we get the following effective mean-field Hamiltonian corresponding to the $i-$th site,
\begin{equation}
h_i=\frac{1}{2}Jr(l_1^2+l_2^2)- Jr(l_1 \cos{\frac{2 \pi i_k}{q}}+l_2\sin{\frac{2 \pi i_k}{q}}).
\label{hi_effH}
\end{equation}
The $q$ possible values of $h_i$ are given by 
\begin{equation}
\epsilon(k)=\frac{1}{2}Jr(l_1^2+l_2^2)- r(l_1 \cos{\frac{2 \pi k}{q}}+l_2\sin{\frac{2 \pi k}{q}}),
\label{engy_effH}
\end{equation}
where $k$ takes any integer value between 1 and $q$. The energy $\epsilon(k)$ does not depend on the site index $i$ for a transitionally invariant system.

\par  In the following, we will calculate the free energy to determine $l_1$ and $l_2$ and to analyze the equilibrium properties from the free energy minima at a given temperature. 

\par The partition function of each spin is given by,
\begin{equation}
z=\sum_{k=1}^{q} e^{-\beta \epsilon(k)}
\label{z_i}
\end{equation}
Since the spins in the mean-field approach are non-interacting, the total partition function of the system is given by $Z={z}^N$. The free energy corresponding to a single spin is 
\begin{equation}
F=-\frac{\ln{z}}{\beta},
\label{F_eng1}
\end{equation}
where $\beta=\frac{1}{k_B T}$ with $k_B$ being the Boltzmann constant and $T$ being the absolute temperature. Written explicitly, we get
\begin{equation}
F=\frac{ Jr (l^2_1 + l^2_2)}{2}-\frac{1}{\beta}\ln\sum_{k=1}^{q} e^{\beta Jr(l_1\cos{\frac{2\pi k}{q}}+l_2 \sin{\frac{2\pi k}{q}})}.
\label{F_eng2}
\end{equation}

\begin{figure*}

{\xincludegraphics[scale=0.28,label=\text{(a)}]{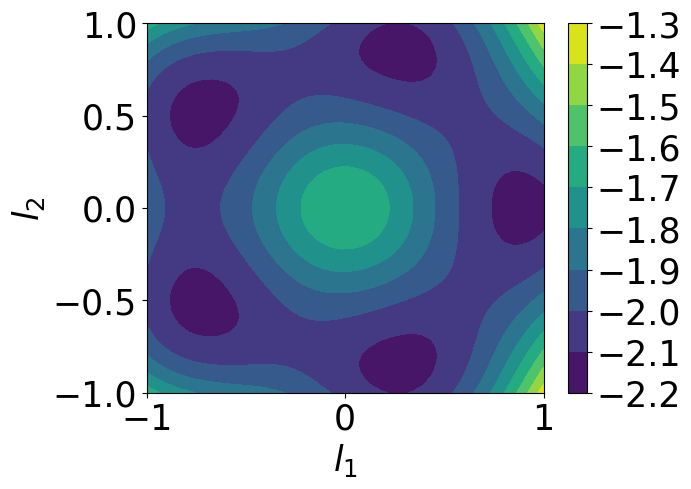}
        \phantomsubcaption\label{1a}}
{\xincludegraphics[scale=0.28,label=\text{(b)}]{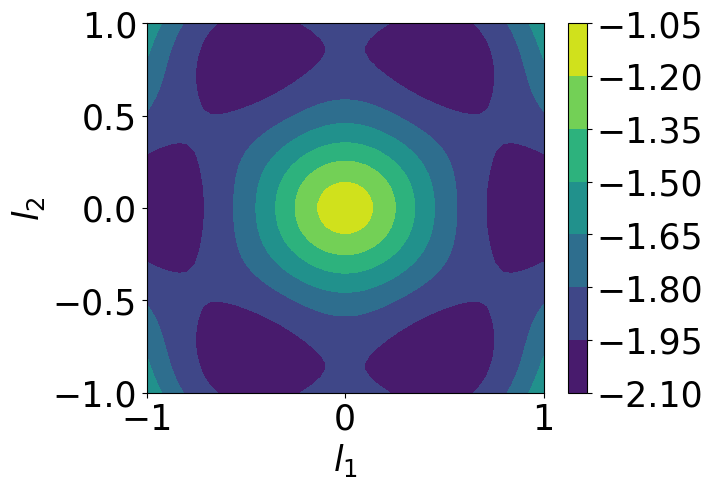}
\phantomsubcaption\label{1b}}
{\xincludegraphics[scale=0.28,label=\text{(c)}]{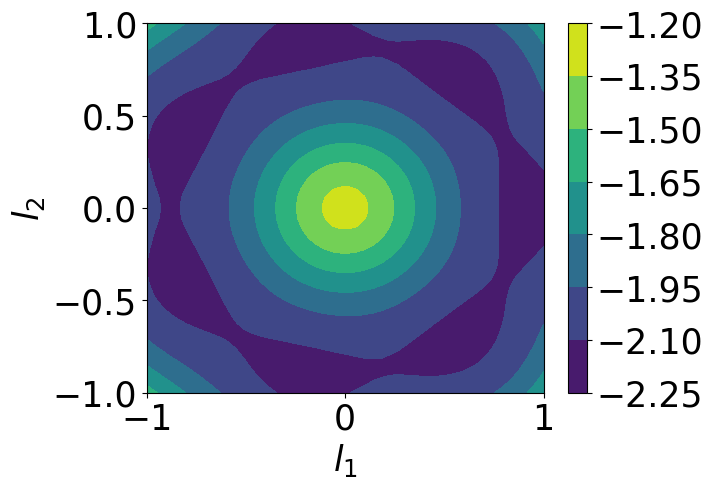}
\phantomsubcaption\label{1c}}

{\xincludegraphics[scale=0.28,label=\text{(d)}]{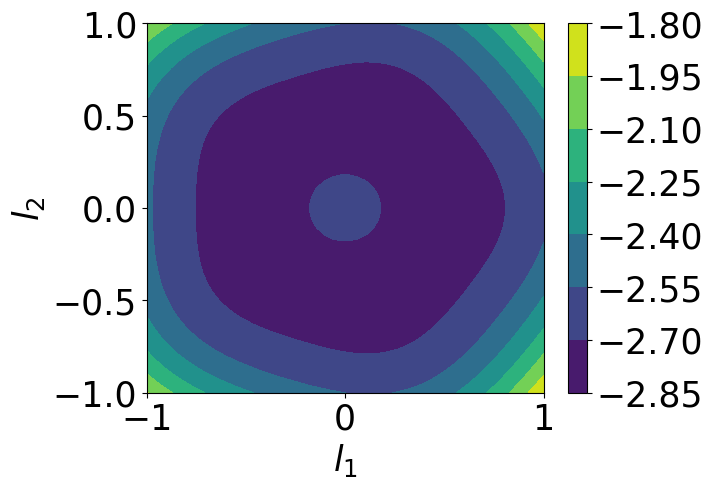}
        \phantomsubcaption\label{1d}}
{\xincludegraphics[scale=0.28,label=\text{(e)}]{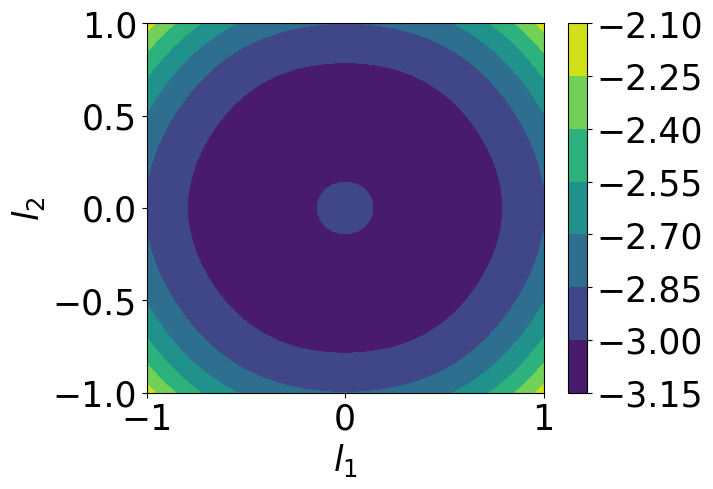}
\phantomsubcaption\label{1e}}
{\xincludegraphics[scale=0.28,label=\text{(f)}]{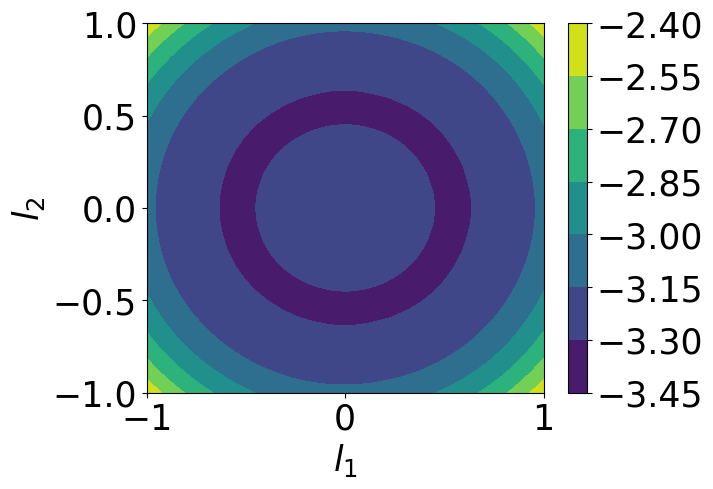}
\phantomsubcaption\label{1f}}

{\xincludegraphics[scale=0.29,label=\text{(g)}]{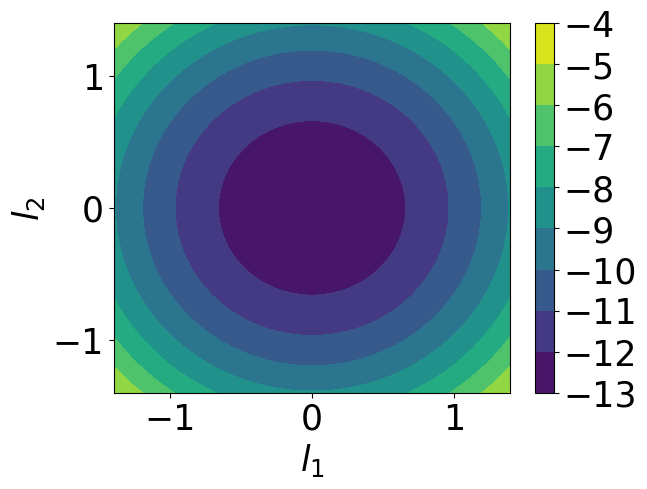}
        \phantomsubcaption\label{1g}}
{\xincludegraphics[scale=0.29,label=\text{(h)}]{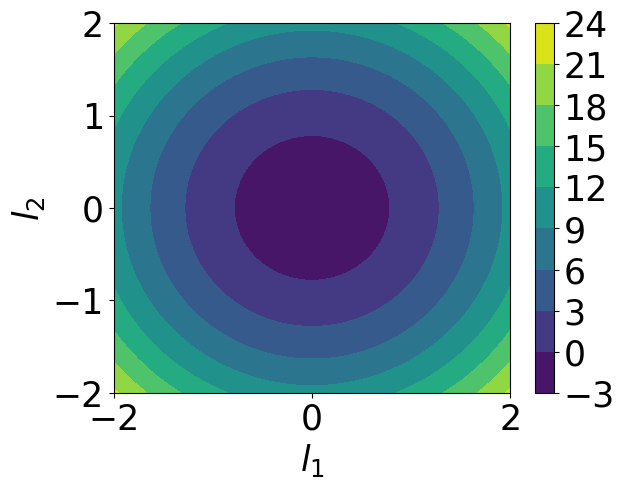}
        \phantomsubcaption\label{1h}}
{\xincludegraphics[scale=0.29,label=\text{(i)}]{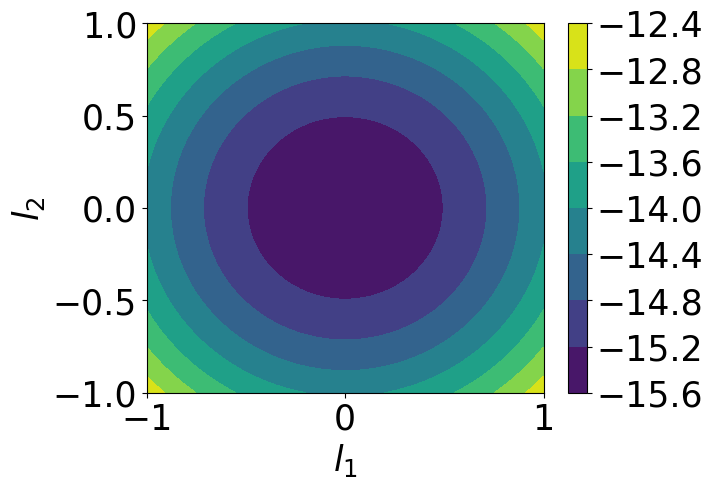}
        \phantomsubcaption\label{1i}}
\caption{Basic mean-field results: free energy plots at low, intermediate and high temperatures are shown in (a)-(c), (d)-(f) and (g)-(i), respectively. Results in the left, middle and right columns are for $q=$ 5, 6 and 7, respectively.}
\label{fig:free_plots}
\end{figure*}

\subsection{Identification of phases from free energy plots}\label{free_plots_0}
The free energy $F$ in Eq. \ref{F_eng2} is a function of $l_1$, $l_2$ and $\beta$. At a given temperature, the equilibrium phase of a system corresponds to the free energy minimum (minima) with respect to $l_1$ and $l_2$. Thus, to first gain insight into the problem, we plot $F$ as a function of $l_1$ and $l_2$ for different fixed temperatures. The plots can be seen in Fig. \ref{fig:free_plots}. 

From the free energy plots one can identify three different phases of the clock model. At high temperatures, we have only one minimum at $l_1=l_2=0$. On the other hand, at low temperatures, we have a local maximum at the origin ($l_1=l_2=0$) but $q$ number of minima around the origin. Interestingly, there is an intermediate temperature range where we get a ring of minima around the origin (in this case, the free energy looks like a $U(1)$ symmetric ``Mexican hat''). 

As mentioned earlier, the parameters $l_1$ and $l_2$, respectively, represent the average values of the $X$ and $Y$ components of a spin. Keeping this in mind, we can have the following interpretations of the above observations. At high temperatures, we have one free energy minimum at the origin ($|l|^2=l_1^2+l_2^2 = 0$). In this phase, an individual spin can orient along any possible direction. This corresponds to the disordered (paramagnetic) phase. As we lower the temperature, we see the minimum turns into a local maximum, and a ring of minima forms around the maxima. This is the emergence of the $U(1)$ symmetry from a $\mathbb{Z}_q$ symmetric model. This is further discussed in Sec. \ref{u1_symm}. Unlike in the paramagnetic phase, in this new phase, an individual spin can orient along a fixed direction ($|l|^2=l_1^2+l_2^2 \neq 0$), but different spins can orient in different directions. In fact, we later show using the higher order MF theory (Sec. \ref{nn_corr}) that the nearest-neighbor correlation takes an intermediate value between 0 and 1 (these two values, respectively, correspond to the disordered phase and ferromagnetic phase). We identify this phase with the bound vortex-antivortex or Berezinskii-Kosterlitz-Thouless (BKT) phase. 

If we further lower the temperature, we get $q$ number of minima around the origin ($l$ takes $q$ possible non-zero (complex) values corresponding to the minima). Although all spins can orient along any of the $q$ possible directions at the low temperature, the spins will spontaneously choose one of them. This phase is identified as the (broken) $\mathbb{Z}_q$ symmetric ordered (ferromagnetic) phase. 

\subsection{Understanding emergence of $U(1)$ symmetry} \label{u1_symm}
As seen in Fig. \ref{fig:free_plots}, also as discussed in some earlier works (see, e.g., Refs. \cite{ortiz2012,vanEnter11,Frohlich81}), there is an emergence of continuous $U(1)$ symmetry in the $\mathbb{Z}_q$ symmetric 2D clock model in the intermediate temperature regime. This emergence can be understood in a straightforward way using mean-field calculations.

To appreciate how a continuous $U(1)$ symmetry emerges in a $\mathbb{Z}_q$ symmetric model, we first express the free energy, found in Eq. \ref{F_eng2}, as a function of parameters $|l|$ and $\theta$, where $l=l_1+i l_2 = |l|e^{i\theta}$. We find that
\begin{equation}
F(|l|,\theta)=\frac{ Jr |l|^2}{2}-\frac{1}{\beta}\ln\sum_{k=1}^{q} e^{\beta Jr|l|\cos{(\frac{2\pi k}{q}-\theta})}.
\label{F_polar}
\end{equation}

In the high or in the intermediate temperature limit (where $\beta$ is small and $|l|<1$), we can effectively expand $e^{\beta Jr|l|\cos{(\frac{2\pi k}{q}-\theta)}}$ in a series. Noting that $\sum_{k=1}^q \cos^n{(\frac{2\pi k}{q}-\theta)}=0 ~\text{or} ~\frac{q}{2^n} C(n,\frac{n}{2})$, respectively, for odd or even $n$ (here $C(n,m)$ denotes a binomial coefficient), we conclude that the free energy $F(|l|,\theta)$ does not depend on $\theta$ in the high or intermediate temperature limits. This result shows that there is an emergence of the continuous $U(1)$ symmetry in the aforementioned temperature limits. This symmetry disappears in the low-temperature limit, and the free energy $F(|l|,\theta)$ starts depending on the parameter $\theta$. As our numerical results show (in Fig. \ref{fig:free_plots}), the free energy will have $q$ number of minima, and the system will recover its $\mathbb{Z}_q$ symmetry (albeit the fact that there will be spontaneous breaking of this discrete symmetry in the low-temperature limit). It is interesting to note here that, although the mean-field theory yields better results for large $q$, our calculations for $q=5$ shows that the model retains its discrete symmetry up to certain extent in the intermediate phase (Fig. \ref{fig:free_plots} d). This finding is consistent with the results reported in Ref. \cite{Baek13}.

To see how the $U(1)$ symmetry disappears and the free energy starts depending on $\theta$, we note that in this low-temperature limit, $\beta$ is very large (and $|l| \approx 1$). In this limit, it is not advisable to do calculations using a series expansion of an exponential function. To understand what happens in this limit, we first replace the sum in Eq. \ref{F_polar} by a more symmetric one in the following way:

\begin{equation}
F(|l|,\delta\theta)=\frac{ Jr |l|^2}{2}-\frac{1}{\beta}\ln\sum_{k=-\frac q2 +1}^{\frac q2} e^{\beta Jr|l|\cos{(\frac{2\pi k}{q}-\delta\theta)}}.
\label{F_polar1}
\end{equation}
Although we assume here $q$ to be even, the results followed are valid for the odd $q$ as well. Since we want to see how the free energy starts depending on the parameter $\theta$, we here assume a small value ($\delta \theta$) for the parameter for mathematical convenience. In the following we would assume $\delta \theta < \frac{2\pi}{q}$. Noting that, in the low temperature limit, $e^{\beta Jr|l|\cos{(\frac{2\pi k}{q}-\delta \theta})}$ is significant only for those values of $k$ for which $|\cos{(\frac{2\pi k}{q}-\delta \theta)}|\approx 1$, we approximate $F(|l|,\delta\theta)$ in the following way:
\begin{equation}
F(|l|,\delta\theta) \approx \frac{ Jr |l|^2}{2}-\frac{1}{\beta}\ln{Q(t, \delta \theta)}, 
\label{F_polar2}
\end{equation}
where $Q(t, \delta \theta)=e^{t\cos{(\frac{2\pi }{q}+\delta\theta})}+e^{t\cos{(\frac{2\pi }{q}-\delta\theta)}}+e^{t\cos{(\delta\theta)}}$ with $t=\beta J r |l|$. Here we only consider the three most significant values for $k$, i.e., $k=0, \pm 1$ in the sum in Eq. \ref{F_polar1}. This approximation is valid as long as we are in a low-temperature regime and $\delta \theta < \frac{2\pi}{q}$. Now if $q$ is large enough so that $\cos{(\frac{2\pi}{q})}\approx 1 - \frac 12 (\frac{2\pi}{q})^2$ and $\sin{(\frac{2\pi}{q})}\approx \frac{2\pi}{q}$, then, 
\begin{equation}
Q(t,\delta\theta)\approx e^{t\cos{\delta\theta}}\left[1+2e^{-\frac 12 (\frac{2\pi}{q})^2t \cos{\delta\theta}}\cosh{\left(\frac{2\pi}{q}t\sin{\delta\theta}\right)}\right]. 
\label{Q_t}
\end{equation}
Now to proceed further, we work in the low-temperature regime where $t(\frac{2\pi}{q})^2 < 1$. This assumption is clearly not valid when $\beta \to \infty$ (or $t\to \infty$), but can be appropriate near the phase transition point at the lower temperature (i.e. near $T_{SSB}$). Noting that $\delta \theta < \frac{2\pi}{q}$, we get the following value of $Q(t,\delta\theta)$ retaining the terms up to $(\delta\theta)^2$:
\begin{equation}
Q(t,\delta\theta)\approx 3 e^t \left[1-\frac 13 t (\frac{2\pi}{q})^2-\frac 12 t (\delta \theta)^2\right]. 
\label{Q1_t}
\end{equation}
The above expression for $Q(t,\delta\theta)$ tells us how the free energy in Eq. \ref{F_polar2} starts depending on the parameter $\theta$ as we lower the temperature, starting from the phase with the emergent $U(1)$ symmetry. The temperature at which this transition happens can be estimated from Eq. \ref{Q1_t}. At the transition temperature (denoted here by $T_{SSB}$), the $\theta$-dependent term becomes significant in comparison to the $\theta$-independent terms in Eq. \ref{Q1_t}, i.e., $1-\frac 13 t (\frac{2\pi}{q})^2=O(t(\delta\theta)^2)$ at the transition temperature. We also note that a physically significant change in $\theta$ is of the order of $\frac{2\pi}{q}$, i.e., we have $\delta\theta = O(\frac 1q)$ in Eq. \ref{Q1_t}. This gives us:  $1-\frac 13 t (\frac{2\pi}{q})^2=O(\frac {t}{q^2})$ at the transition temperature. Solving this we get that $t=O(q^2)$ or $T_{SSB}=O(\frac{1}{q^2})$. 

\subsection{Nature of two phase transitions}
The most of the earlier works predicted that the transition between the high temperature paramagnetic phase and the intermediate phase is of the BKT type \cite{Li2020,Hong2020,ortiz2012,Surungan2019,Kumano13}. Our numerical results also support this finding. As discussed in the last subsection and also apparent from Fig. \ref{fig:free_plots}, there is an intermediate phase with an emergent $U(1)$ symmetry. Now, as we lower the temperature, the system goes from the paramagnetic phase to the intermediate phase with the emergent $U(1)$ symmetry, without breaking any continuous symmetry. This indicates that the corresponding phase transition is of the BKT type. The other transition between the ferromagnetic phase and the $U(1)$ symmetric BKT phase needs a more careful analysis; we discuss this in the following.

Although there is a general consensus on the nature of the phase transition at the higher temperature, the controversy about the nature of the phase transition at the lower temperature is still unresolved. Many recent studies have found that this transition between the ferromagnetic and BKT phases is of the type BKT \cite{Li2020,ortiz2012,Surungan2019}. There are also some studies where the BKT nature of this transition is ruled out \cite{Lapilli06,Hwang09}. In many works, the BKT nature of the transition at the lower temperature is just assumed or conjectured \cite{Hong2020,negrete2021short}.

In such a doubtful situation, one can go back to the basic results related to the phase transitions for a better understanding of the nature of the phase transition at the lower temperature. We note that (a) the actual BKT transition is not a (spontaneous) symmetry-breaking transition, and (b) for the clock model, the phase transition at the lower temperature breaks the symmetry of the system (from emergent $U(1)$ to broken $\mathbb{Z}_q$). These two observations firmly suggest that the transition at the lower temperature is not a BKT type transition and it must be one type of the spontaneous symmetry-breaking (SSB) transition. Another supporting fact in favor of the SSB transition is that one can define here a local order parameter (average magnetization or spin expectation value) whose value will be zero in the critical intermediate phase (the BKT phase) and will be non-zero in the low-temperature ordered ferromagnetic phase. We note that one can not have any local order parameter for a BKT transition.     

Following the above arguments, if one accepts that the transition at the lower temperature is an SSB transition, one can still debate the possible order of the transition (since any finite-order transition can, in principle, be an SSB transition). Earlier numerical works ruled out that the transition is not a first- or second-order transition \cite{Hong2020,Borisenko11}. This leaves us with the possibility that the phase transition at the lower temperature is a large-order phase transition (the order is supposed to be greater than two but less than infinity). Here, our mean-field theory can give some supporting evidence for the fact that the transition could possibly be a large-order transition. We see from Fig. \ref{fig:free_plots} or Fig. \ref{fig:free_plots1} of our manuscript that, in the low temperature, the free energy has $q$ minima and the same number of local maxima in between them. That indicates that the free energy can be written (\`{a} la Ginzburg-Landau) as an $n$-degree polynomial of the order parameter $\tilde{\theta}$ where $n$ can be taken to be $2q+2$: 
\begin{equation}
F(T,\tilde{l},\tilde{\theta}) = F(T,\tilde{l})+\sum_{i=0}^q C_i(T,\tilde{l})\tilde{\theta}^{2i+2}
\label{GL_FE}
\end{equation}
with $-\pi < \tilde{\theta} \le \pi$. We note that the free energy depends on two local order parameters: $\tilde{\theta}$ which is a local average of the angle $\theta$ and $\tilde{l}$ which is a local average of $|l|$. In Eq. \ref{GL_FE}, the constants $C_i$ are generally temperature dependent and also depend on $\tilde{l}$. The order parameter $\tilde{l}$ itself depends on the temperature: it is zero when $T>T_{SSB}$ but quickly reaches the value 1 as the temperature goes below $T_{SSB}$ and the system moves into the ferromagnetic phase. The special form of $F$ in Eq. \ref{GL_FE} is taken so that $F(-\tilde{\theta}) = F(\tilde{\theta})$, as expected for a system without a magnetic field. The degree of the polynomial in $\tilde{\theta}$ is taken to be $2q+2$ so that the solutions of the equation $\partial F/\partial\tilde{\theta} = 0$ correspond to the desired number of minima and maxima.

We note that, since there is always a minimum corresponding to $\tilde{\theta}=0$ (irrespective of whether $q$ is even or odd), it is enough to analyze $F(T,\tilde{l},0) = F(T,\tilde{l})$ to know the order of the phase transition. To this effort, one needs to know how the free energy $F(T,\tilde{l})$ can be expressed as a polynomial in $\tilde{l}$ near the phase transition point. Unfortunately, this issue can not be addressed satisfactorily within the mean-field approach presented in this paper. On can simply guess, say from Eq. \ref{Q1_t}, that if the free energy is a large-degree polynomial in $\tilde{\theta}$, then it would probably be a large-degree polynomial in $\tilde{l}$ also. We know from the theory of higher-order (greater than second order) phase transition that one requires a large-degree polynomial in the order parameter to describe the transition \cite{Kumar03,Ekuma12}. This leads to our conjecture on the nature of the phase transition at the lower temperature in the 2D clock model for large $q$.  

Although the mean-field theory is not adequate to determine the order of transition at $T_{SSB}$, we discuss now a possible way to determine the order of transition. The free energy minimum (with respect to $\tilde{l}$ for $\tilde{\theta}=0$) near the transition point can be written as $-F(T)=A\left(1-\frac{T}{T_{SSB}}\right)^{p'}$, where $A$ is a (positive) constant and $T\le T_{SSB}$. We note that, in reality, $p'$ may not be an integer; the actual order of transition (an integer) is given by $p=p'+\mu$ with $\mu$ being a small correction factor. Noting that the entropy $S(T)=-\frac{\partial F}{\partial T}$ and the heat capacity $C(T)=-T \frac{\partial^2F}{\partial T^2}$, we have 
\begin{equation}
p'=1+\lim_{T\to T_{SSB}}\left(1-\frac{T}{T_{SSB}}\right) \frac{C(T)}{S(T)}.
\label{order_tssb}
\end{equation}

To estimate $p'$ of the above equation, consider that $\{\varepsilon_i\}$ is the set of energy values of the system and $p_i=\frac{1}{Z}e^{-\beta \varepsilon_i}$ is the corresponding Boltzmann probability, where the partition function $Z=\sum_i e^{-\beta \varepsilon_i}$ with $\beta$ being the inverse temperature. The internal energy of the system is given by $U(T) = \sum_i p_i\varepsilon_i = -\frac{\partial \ln Z}{\partial \beta}$. One can now calculate the heat capacity using: $C(T) = \frac{\partial U}{\partial T}$. Similarly, the entropy can be calculated using: $S(T) = - \sum_i p_i \ln p_i$. Thus we see that, if we can simulate the set of energy values $\{\varepsilon_i\}$ and corresponding probability distribution $\{p_i\}$, we can estimate $p'$ of Eq. \ref{order_tssb} by evaluating $C(T)/S(T)$. As mentioned earlier, the mean-field theory will not be adequate for such estimation; the Monte Carlo or other suitable numerical method may be employed for such a purpose.

\subsection{Calculation of $T_{BKT}$}
In this part we estimate $T_{BKT}$, the temperature at which transition happens between the BKT and the paramagnetic phases. This transition corresponds to the free energy minimum at the origin ($l_1=l_2=0$) turning into a local maximum. Thus, we will get an inflection of the free energy $F(l_1,l_2)$ at the temperature $T_{BKT}$.

\par In practice, we get the transition temperature from either of the equations 
$\frac{\partial^{2}F}{\partial l_1^{2}}|_{(0,0)}=0$
or $\frac{\partial^{2}F}{\partial l_2^{2}}|_{(0,0)}=0$. From Eq. \ref{F_eng2}, we get

\begin{widetext}
\[ 
\pdv[order={2}]{F}{l_1}_{(0,0)}^{}
=Jr-\frac{{(Jr)}^2 \beta \sum_{j=1}^{q} 1 \sum_{j=1}^{q} \cos^2{\frac{2\pi j}{q}}}{{(\sum_{j=1}^{q} 1)}^2} + \frac{\beta {(Jr)}^2{(\sum_{j=1}^{q} \cos{\frac{2\pi j}{q})}}^2}{{(\sum_{j=1}^{q} 1)}^2} =0, ~\text{i.e.,}
\]
\begin{equation}
Jr-\frac{{(Jr)}^2 \beta \sum_{j=1}^{q} \cos^2{\frac{2\pi j}{q}}}{q} + \frac{\beta {(Jr)}^2{(\sum_{j=1}^{q} \cos{\frac{2\pi j}{q})}}^2}{q^2} =0
\label{Tbkt1}
\end{equation}
\end{widetext}

We now note the following trigonometric identities,  
\[\sum_{j=1}^{q} \cos{\frac{2\pi j}{q}}=0, ~ \text{and}\]
\[\sum_{j=1}^{q} \cos^2{\frac{2\pi j}{q}}=\frac{\sum_{j=1}^{q} (1+\cos{\frac{4\pi j}{q}})}{2}
=\frac{q}{2}.\]
We get the second identity by using $\sum_{j=1}^{q} cos{\frac{4\pi j}{q}}=0$.

\noindent Thus, we get from Eq. \ref{Tbkt1} 
\[Jr -\frac{(Jr)^2 \beta}{2} =0.\] Noting that, $\frac{1}{\beta}=k_B T_{BKT}$, we finally get
\begin{equation}
T_{BKT}=\frac{Jr}{2k_B}
\label{Tbkt2}
\end{equation}
For a two-dimensional square lattice $r=4$, hence, $T_{BKT}=\frac{2J}{k_B}$. We note here that the BKT transition temperature is independent of $q$. We again estimate this temperature using the first-order MF theory in Sec. \ref{sec3}. 

\subsection{Calculation of $T_{SSB}$}
 As we already have seen in Sec. \ref{u1_symm}, there is a transition temperature at which the emergent $U(1)$ symmetry breaks. This temperature, denoted here by $T_{SSB}$, decreases quadratically over the variable $q$, i.e., $T_{SSB}=O(\frac{1}{q^2})$. This can also be understood in the following way. At very low temperatures, all the spins align in a particular direction. This ferromagnetic order is broken due to the low-lying excitations as we increase the temperature. The temperature of transition ($T_{SSB}$) corresponds to the minimum energy required to break the ferromagnetic order.  We note that the larger the value of $q$, the less energy we need to break the ferromagnetic order. This shows that $T_{SSB}$ depends on $q$, decreasing as $q$ increases. To know how $T_{SSB}$ functionally depends on $q$, we must first calculate the energy required for a spin to make the smallest possible angle with its original direction in the ferromagnetic phase. Eq. \ref{H_clc} shows that this energy is proportional to $1-cos(2\pi/q)$, i.e., proportional to $1/q^2$ for large $q$. Thus, we infer that $T_{SSB} \propto 1/q^2$.
 
\par In the following we go on to estimate $T_{SSB}$ using yet another approach. We recall that, see Fig. \ref{fig:free_plots}, at low temperature (when the system is in ferromagnetic phase), all the spins spontaneously choose one of the $q$ free energy minima points. As we increase temperature, this ferromagnetic order will be broken when spins move from one minimum to the next. We note that while moving from one minimum to the next, the system needs to cross a barrier (local maximum). So we set the following criterion to estimate $T_{SSB}$: this is the temperature at which thermal energy at any of the minima equals the barrier height between two consecutive minima. 

The order of the thermal energy is calculated from $TS=U-F$; for this purpose, we first separately calculate internal energy ($U$) and the free energy $F$) at one of the $q$ global minima. While calculating the energy barrier, we calculate the difference between one of the global minima and the local maximum between two consecutive global minima.
 
\par Calculating the energy barrier requires different strategies for even and odd values of $q$. We describe them separately below. 

\subsubsection{Calculation of $T_{SSB}$ when $q$ is odd}
\begin{figure*}
    \centering
       \hspace{-0.1cm} {\xincludegraphics[scale=0.25,label=\text{(a)}]{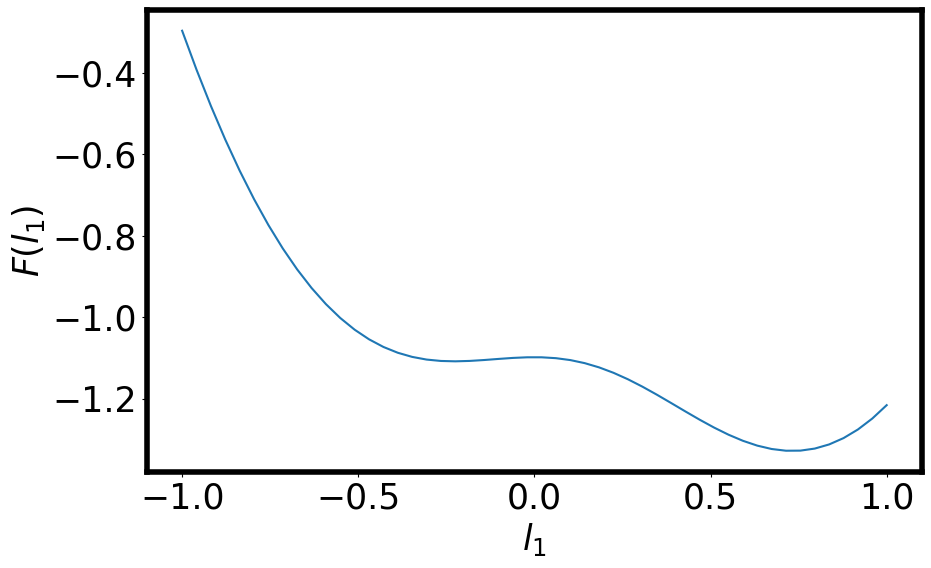}
        \phantomsubcaption\label{fig:q=3 at beta=0.8}}
        \hspace{-0.05cm}
    {\xincludegraphics[scale=0.25,label=\text{(b)}]{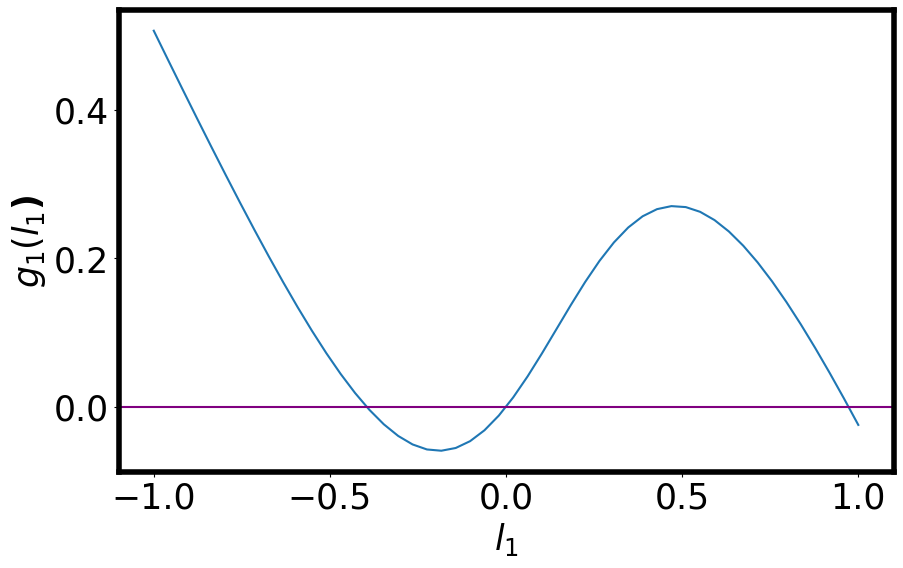}
        \phantomsubcaption\label{fig:q=3,beta=0.8}}
    {\xincludegraphics[scale=0.25,label=\text(c)]{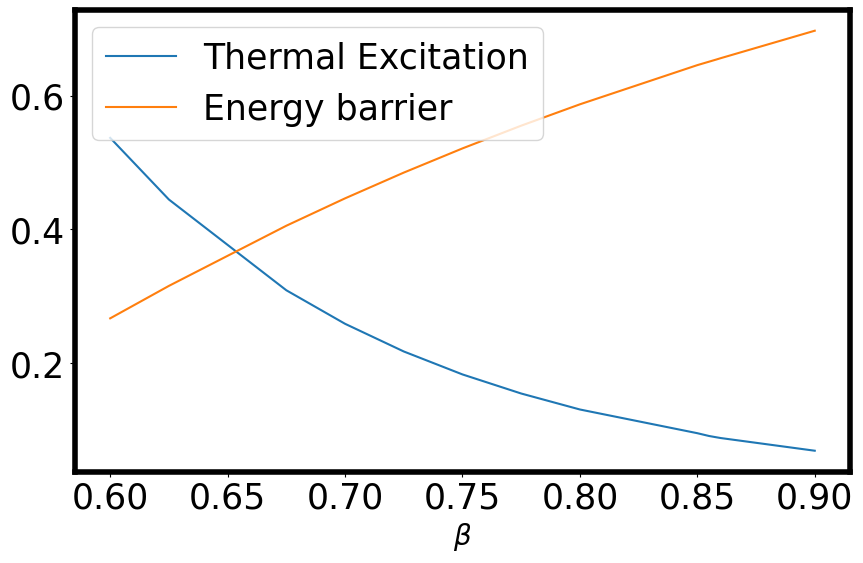}
        \phantomsubcaption\label{fig:T3}}
        \renewcommand{\thefigure}{2}
    \caption{
        Plots of (a) $ F(l_1,0) $ vs. $l_1$ and (b) $ g(l_1) $ vs. $l_1$ for $q=3$ clock model with $\beta=0.82$. (c) The thermal energy and the energy barrier are plotted as function of $\beta$ for $q=3$. $T_{SSB}$ is estimated from where the two lines cross.}
    \label{Tssb_q3}
\end{figure*}
For odd $q$, in the low-temperature regime (Fig. \ref{fig:free_plots}), we have one global minimum on the positive $l_1$-axis and one local maximum (between two consecutive global minima) on the negative $l_1$-axis. For illustration, we plot the low-temperature free energy as a function of $l_1$ keeping $l_2=0$ for the $q=3$ model, as shown Fig. \ref{Tssb_q3}(a)). In this plot, we see a global minimum and a local minimum, which is actually the local maximum while going from one to the next global minimum. We can calculate their positions along the $l_1$-axis by taking $\pdv{F(l_1,0)}{l_l}=0$, where $F(l_1,0)$ represents the free energy as function of $l_1$ keeping $l_2=0$.

\par For $q=3$, according to Eq. \ref{F_eng2}, the free energy function $ F(l_1,0)$ is
 \begin{equation}
F(l_1,0)=\frac{ Jl^2_1 r}{2}-\frac{\ln\sum_{i=1}^{3}e^{{\beta {Jr(l_1\cos{\frac{2\pi i}{q}})}}}}{\beta}. 
\label{F_q3}
\end{equation}
Taking the partial derivative of $F(l_1,0)$ w.r.t $l_1$, and equating it to zero, we get
\begin{equation} 
    l_1=\frac{\sum_{i=1}^{3} \cos{\frac{2\pi i}{3}} e^{\beta l_1 Jr \cos{\frac{2\pi i}{3}}}}{\sum_{i=1}^{3}e^{\beta l_1 Jr \cos{\frac{2\pi i}{3}}}}.
\label{l1_min1}
\end{equation}
Considering $r=4$ and $J=1$, we obtain
\begin{equation}
    l_1=\frac{e^{4 \beta l_1}-e^{-2 \beta l_1}}{e^{ 4 \beta l_1 }+2e^{-2\beta l_1 }}.
\label{l1_min2}
\end{equation}

This is a transcendental equation that can not be solved directly. We need to do some numerical calculations to solve this equation. Let
\begin{equation}
   g(l_1)= \frac{e^{4 \beta l_1}-e^{-2 \beta l_1}}{e^{ 4 \beta l_1 }+2e^{-2\beta l_1 }}-l_1.
\end{equation}
For a given $\beta$, the values of $l_1$ corresponding to the zeros of the function $g(l_1)$ are the solutions of Eq. \ref{l1_min2}. These solutions can easily be found from the plot of $g(l_1)$, as can be seen in Fig. \ref{Tssb_q3}(b). We clearly see that there are three solutions:  one corresponds to the global minimum ($l_1>0$), the next one corresponds to the local maximum ($l_1=0$), and the last one corresponds to the local maximum between two consecutive minima ($l_1<0)$. Now, the energy barrier can be found by taking the difference between the global minimum and local maximum calculated from $F(l_1,0)$ using appropriate $l_1$ values (solutions of Eq. \ref{l1_min2}).
The energy barrier as a function of $\beta$ can be found in Fig. \ref{Tssb_q3}(c).

Now, for the calculation of thermal energy, we first calculate the internal energy of the spin by using the relation
\[U=\frac{\sum_{k=1}^{3} \epsilon(k) e^{-\beta \epsilon(k)}}{\sum_{k=1}^{3}e^{-\beta \epsilon(k)}},\]
where the expression of energy $\epsilon(k)$ can be found in Eq. \ref{engy_effH} with $q=3$. Written explicitly, we have
\begin{equation}
    U=\frac{2(2l^2_1+2l_1)e^{-2l_1 \beta}+(2l^2_1-4l_1) e^{4\beta l_1}}{2e^{-2\beta l_1}+ e^{4\beta l_1}}.
\label{U_eng}
\end{equation}

The order of thermal energy ($U-F$) at a global minimum is calculated using Eqs. \ref{F_q3} and \ref{U_eng} with the appropriate $l_1$ value coming from solving Eq. \ref{l1_min2}. The thermal energy as a function of $\beta$ can be found in Fig. \ref{Tssb_q3}(c). From this plot, we can easily check the temperature at which the thermal energy and the energy barrier match. We find that around $\beta=0.65$, they are equal; thus, for $q=3$, the symmetry-breaking phase transition temperature is $T_{SSB}=\frac{1}{0.65k_B}=\frac{1.538}{k_B}$. In the same way, we can calculate symmetry-breaking phase transition temperature for other odd $q$ values. For $q=5$ and $7$, calculated $T_{SSB}$ values can be found in Table \ref{Tssb}.

\subsubsection{Calculation of $T_{SSB}$ when $q$ is even}
\begin{figure*}
    \centering
        {\xincludegraphics[scale=0.26,label=\text{(a)}]{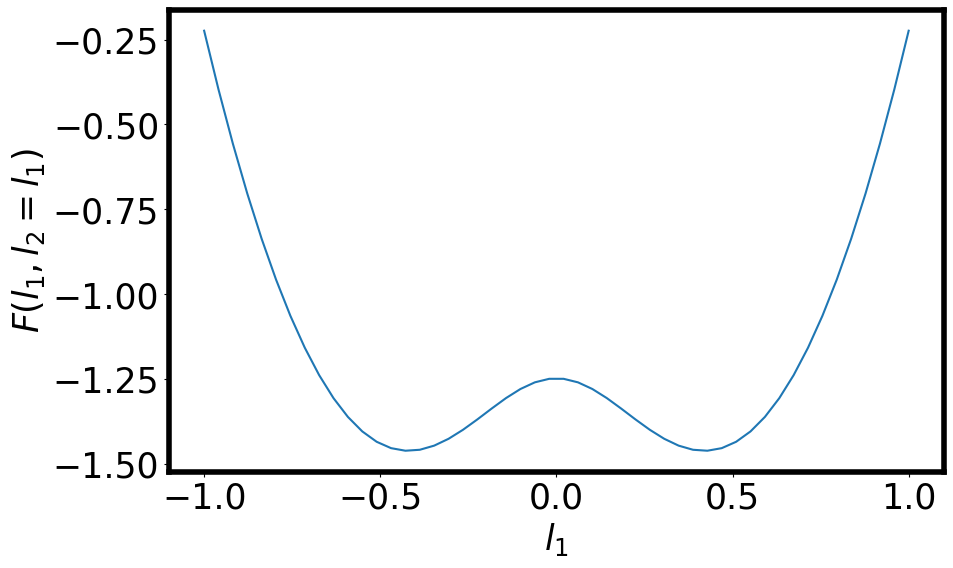}
        \phantomsubcaption\label{F_min1}}
        {\xincludegraphics[scale=0.26,label=\text{(b)}]{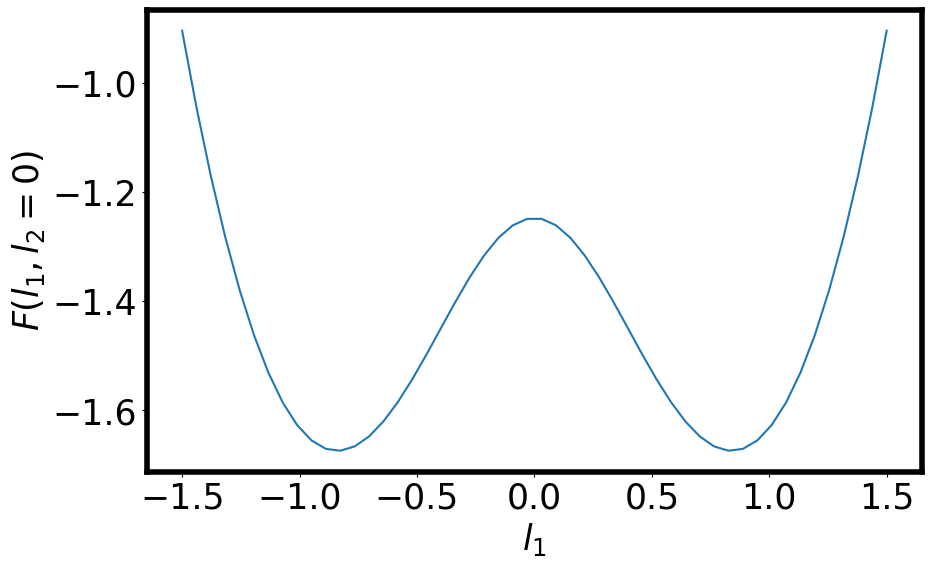}
        \phantomsubcaption\label{F_min2}}
        
    \centering
    {\xincludegraphics[scale=0.26,label=\text{(c)}]{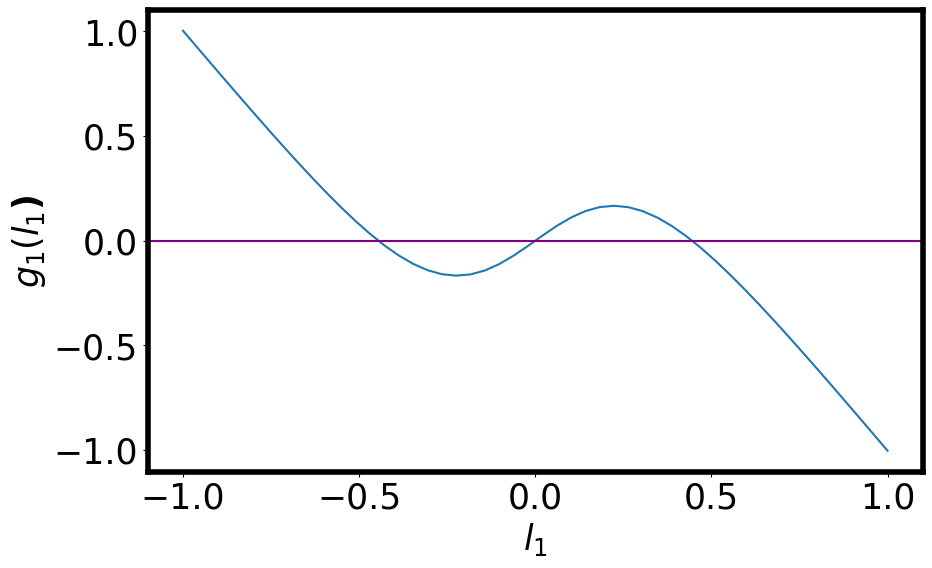}
        \phantomsubcaption\label{fig:subim1}}
        {\xincludegraphics[scale=0.26,label=\text{(d)}]{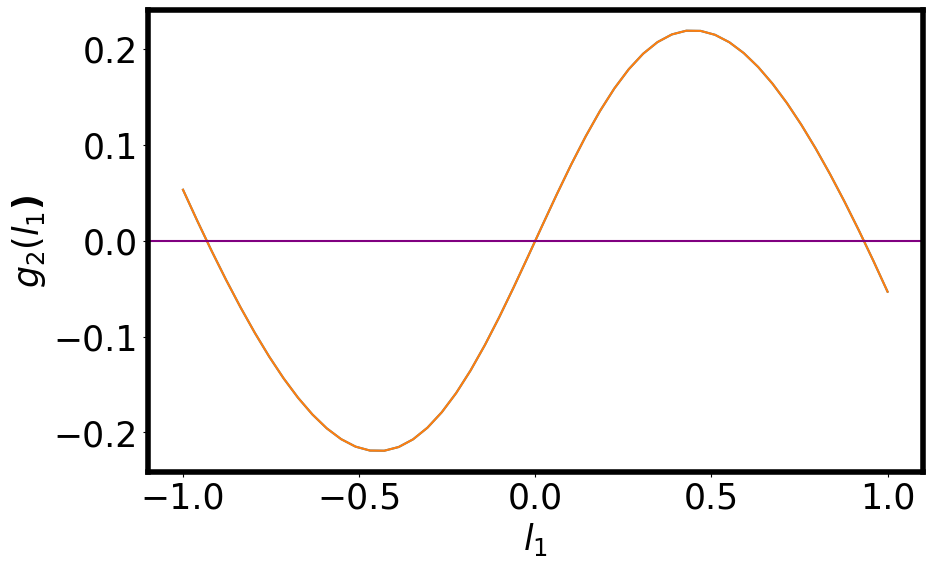}
        \phantomsubcaption\label{fig:subim2}}
    \caption{Plots of (a) $ F(l_1,l_2 = l_1) $ vs. $l_1$, (b) $F(l_1,0)$ vs. $l_1$, (c) $g_1(l_1)$ vs. $l_1$ and (d) $g_2(l_1)$ vs. $l_1$ for $q=4$ clock model with $\beta=0.82$.}
    \label{Tssb_q4}
\end{figure*}

\begin{figure}[h]
    \centering
     \includegraphics[width=7cm, height=4.5cm]{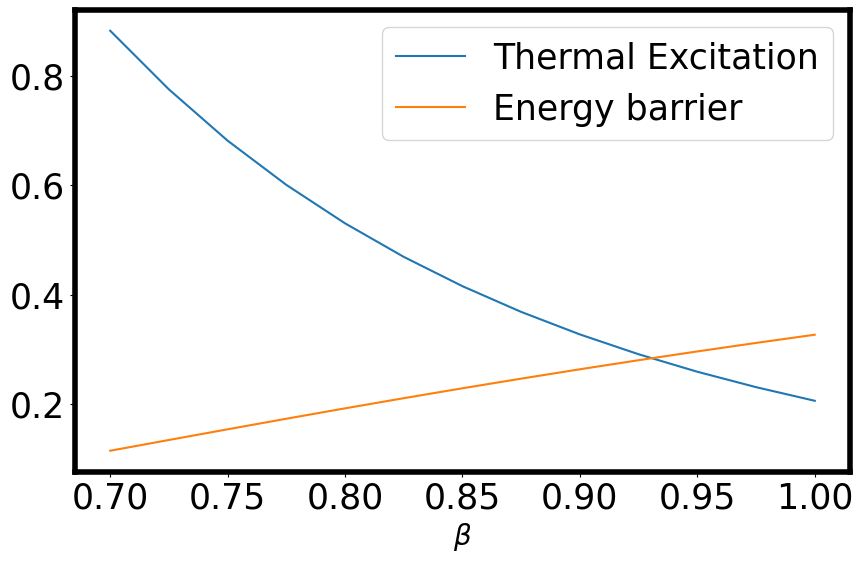}
   \caption{The thermal energy and the energy barrier are plotted as function of $\beta$ for $q=4$. $T_{SSB}$ is estimated from where the two lines cross.}
         \label{fig:q=4}
\end{figure}

While calculating $T_{SSB}$ for the $q-$state clock model with even $q$, the trick used for odd $q$ will not work. If we plot the function $F(l_1,0)$ with $l_1$ or plot $F(0,l_2)$ with $l_2$, we will get two global minima, as can be seen in Fig. \ref{Tssb_q4}(b). However, we need the global minimum and the local maximum (between two successive global minima) to calculate the energy barrier between two successive global minima. To get the local maximum, we need to analyze the free energy along the line joining the origin and the midpoint between two consecutive global minima. For convenience, we can take the line joining the origin and the midpoint between the first two global minima (the first one being on the positive $l_1$-axis). The equation for this line is $l_2=l_1\tan{\frac{\pi}{q}}$. Along this line, the $F(l_1,l_2)$ plot can be seen in Fig. \ref{Tssb_q4}(a). Along this line, we can get the local maximum by solving the equation $\pdv{F}{l_1}=0$. Replacing $l_2$ by $l_1\tan{\frac{\pi}{q}}$ in the expression of $F(l_1,l_2)$, then following the procedure described for odd $q$, we get the following transcendental equation for $l_1$ (taking $q=4$ and $J=1$):
 \begin{equation}
  l_1=\frac{e^{4\beta l_1}-e^{-4\beta l_1}}{2(e^{4\beta l_1}+e^{-4\beta l_1})}.
  \label{q4lm_l1}
 \end{equation}
To solve this equation for $l_1$, we define the following function,
\begin{equation}
g_1(l_1)=\frac{e^{4\beta l_1}-e^{-4\beta l_1}}{2(e^{4\beta l_1}+e^{-4\beta l_1})}-l_1.
\label{q4lm_g1}
\end{equation}
The roots of the equation $g_1(l_1)=0$ are the solutions of Eq. \ref{q4lm_l1}. As can be seen from Fig. \ref{Tssb_q4}(c), there are three solutions. Two solutions for which $l_1\ne 0$ correspond to the local maxima (between global minima).  

There are two global minima on the $l_1$-axis, as can be seen in Fig. \ref{Tssb_q4}(b); we can get them by solving $\pdv{F(l_1,0)}{l_1}=0$. This leads to the following transcendental equation for $l_1$ (again taking $q=4$ and $J=1$): 
\begin{equation} 
l_1=\frac{e^{4\beta l_1}-e^{-4\beta l_1}}{2+(e^{4\beta l_1}+e^{-4\beta l_1})}
\label{q4gm_l1}
\end{equation}
We now define the following function $g_2(l_1)$ to find the solutions of the above equation,
\begin{equation} 
g_2(l_1)=l_1-\frac{e^{4\beta l_1}-e^{-4\beta l_1}}{2+(e^{4\beta l_1}+e^{-4\beta l_1})}.
\end{equation}
We see from Fig. \ref{Tssb_q4}(d) that there are three solutions, two of them for which $l_1\ne 0$ corresponds to the global minima. 
The energy barrier can be found by taking the difference between the global minimum and local maximum of the free energy as calculated above. The energy barrier as a function of $\beta$ can be seen in Fig. \ref{fig:q=4}.

Next, to estimate the thermal energy ($U-F$) at the global minimum, we need to calculate the internal energy $U$. The expression of $U$, in this case, is given by
\begin{equation}
    U=\frac{4l^2_1+(2l^2_1+4l_1)e^{-4\beta l_1}+(2l^2_1-4l_1)e^{4\beta l_1}}{2+e^{-4\beta l_1}+e^{4\beta l_1}},
\end{equation}
where $l_1$ corresponds to a global minimum. The details of how the internal energy is calculated are already discussed for the case of odd $q$. The thermal energy at a global minimum as a function of $\beta$ can be found in Fig. \ref{fig:q=4}. From this figure, we can easily check at which temperature the thermal energy matches with the energy barrier; we see that, for $q=4$, two quantities match near $\beta=0.93$. We, therefore, estimate that $T_{SSB}=1.075/k_B$. We also calculated $T_{SSB}$ for $q=6$, which can be found in Table \ref{Tssb}.

\begin{figure}[h]
    \centering
     \includegraphics[width=8cm, height=5cm]{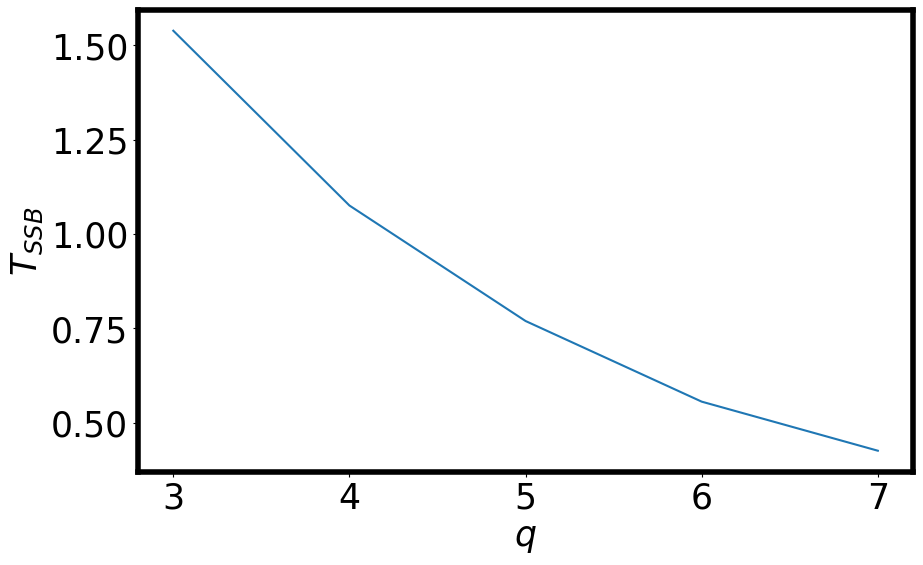}
   \caption{\centering Plot of $T_{SSB}$ vs. $q$}
         \label{Tssb_q}
\end{figure}

From this Table \ref{Tssb}, we can see that the spontaneous symmetry-breaking phase transition temperature ($T_{SSB}$) decreases with an increase in $q$; this is also evident in Fig. \ref{Tssb_q}. Although these results help us understand how $T_{SSB}$ changes with $q$, they do not reveal the exact functional dependence of the transition temperature on $q$. For a better understanding of the functional dependence, in the following, we use our mean-field theory to analytically estimate the value of $T_{SSB}$ in the large-$q$ limit. 

\begin{table}
\begin{ruledtabular}
\begin{tabular}{c c c c }
& $q$-value & $T_{SSB}$\\
\hline
& 3 & $1.538/k_B$\\
& 4 & $1.075/k_B$\\
& 5 & $0.7692/k_B$ \\
& 6 & $0.556/k_B$ \\
& 7 & $0.4225/k_B$\\
\end{tabular}
\end{ruledtabular}
\caption{Numerically calculated $T_{SSB}$ values for different $q$ values}
\label{Tssb}
\end{table}

\subsection{Analytical calculation of $T_{SSB}$ in the large-$q$ limit}
In this part, we first calculate the energy barrier, and then we calculate the thermal energy to estimate $T_{SSB}$ in the large-$q$ limit.
\subsubsection{Calculation of energy barrier}
For the large-$q$ limit, we can convert the summation appearing in the free energy expression, as seen in Eq.\ref{F_eng2}, into an integration. Consequently, we have the following expression of the free energy, 
\begin{equation}
F=\frac{ J (l^2_1 + l^2_2) r}{2}-\frac{1}{\beta}\ln\int_{1}^{q}e^{{\beta {Jr(l_1\cos{\frac{2\pi j}{q}}+l_2 \sin{\frac{2\pi j}{q}})}}}dj.
\label{F_eng3}
\end{equation}
However, performing this integration is not straightforward. To go further, we replace the integrand function with a judiciously chosen simpler function. We first convert the integration limits to a more symmetric ones; accordingly, 
\begin{equation}
F=\frac{ J (l^2_1 + l^2_2) r}{2}-\frac{1}{\beta}\ln\int_{-\frac q2+1}^{\frac q2}e^{{\beta {Jr(l_1\cos{\frac{2\pi j}{q}}+l_2 \sin{\frac{2\pi j}{q}})}}}dj.
\label{F_eng4}
\end{equation}
We only consider the even $q$ for the current calculations. If we are interested in calculating the global minima on the $l_1$-axis, we have to deal with the integrand function $e^{\beta {Jrl_1\cos{\frac{2\pi j}{q}}}}$. While replacing this function with a simpler function, we choose a function so that the extrema points of the original and the new function match. We note that the original function has a maximum at $j=0$ and the function effectively vanishes when $j>\frac q4$ or $j<-\frac q4$, since $\cos{\frac{2\pi j}{q}}$ picks up a negative sign in the ranges. With these properties in mind, we now replace this integrand function with the following approximate function:
\[e^{{\beta {Jrl_1\cos{\frac{2\pi j}{q}}}}}\to (1-c j^2)e^{\beta {Jrl_1}}.\]
The value of $c$ is determined by the fact that the original function effectively vanishes at $j=\pm \frac q4$. This gives us $c=(\frac{4}{q})^2$. It may be mentioned here that the original and the approximate functions have the same maximum, and they effectively vanish at the same points. It is also worth commenting that the integration over the new approximate function should be performed only from $-\frac q4$ to $\frac q4$. With this judiciously chosen approximate function, it is now easy to estimate the value of the integral:
\[{\int_{-\frac q2+1}^{\frac q2}e^{{\beta {Jr(l_1\cos{\frac{2\pi j}{q}})}}}dj}\to{\int_{-\frac{q}{4}}^{\frac{q}{4}}(1-{(\frac{4}{q})}^2 j^2)e^{\beta {Jrl_1}} dj} \]
\[=\frac{q}{3} e^{J r \beta l_1}.\]
Using this result, the expression for the free energy $F$ along the $l_1$-axis can be written as, 
\begin{equation}
F=\frac{ J l^2_1 r}{2}- Jrl_1-\frac{\ln(\frac{q}{3})}{\beta}.
\end{equation}
Now the global minimum can be found from $\pdv{F}{l_1}=0$. This gives us $l_1=1$ at the global minimum. So, the free energy at the global minimum is,
\begin{equation}
F_G=-\frac{Jr}{2}-\frac{1}{\beta}\ln(\frac q3).
\label{F_G}
\end{equation}

For the calculation of the energy barrier, we next calculate the local maximum (between consecutive global minima). We note that if we connect by straight lines all the global minima to the origin ($l_1$=0, $l_2$=0), the angle between consecutive lines would be $\frac{2\pi}{q}$. Let us consider that the coordinate of the local maximum, between the first two global minima, is ($l_1$, $l_2$) and the coordinate of the global minimum, which is on the $l_1$-axis, is given by $(l,0)$. Some trigonometric considerations show that $l_1= l \cos^2{\frac{\pi}{q}}$ and $l_2=l \cos{\frac{\pi}{q}} \sin{\frac{\pi}{q}}$. 
Then, the expression for free energy, as seen in Eq. \ref{F_eng4}, at the local maximum will take the following form:
\begin{equation}
F=\frac{ J r l^2 \cos^2{\frac{\pi}{q}}}{2}-\frac{1}{\beta}\ln\int_{-\frac q2+1}^{\frac q2} e^{{\beta {Jrl \cos{\frac{\pi}{q}} \cos{\frac{(2\pi j- \pi)}{q}}}}}dj.
\label{F_eng5}
\end{equation}
Once again, we will replace the integrand function with a simpler function with the same extreme points. We note that the function $e^{\beta {Jrl \cos{\frac{\pi}{q}}\cos{\frac{(2\pi j- \pi)}{q}}}}$ effectively vanishes when $j<\frac{-q}{4}+\frac{1}{2}$ or $j>\frac{q}{4}+\frac{1}{2}$, and has a maximum at $j=\frac{1}{2}$. But $j$ is always an integer, so we have to consider the nearest integer of $\frac{1}{2}$, which are 0 and 1, while approximating the function; accordingly, the new function will be defined in two regions: one is $(\frac{-q}{4}+\frac{1}{2})$ to 0, and the other one is from 1 to $(\frac{q}{4}+\frac{1}{2})$.  

In the first interval, i.e. from $(\frac{-q}{4}+\frac{1}{2})$ to 0, we can approximate the function as $(1-c j^2) e^{(J r \beta l \cos^2{\frac{\pi}{q}})}$. Both the original function and this approximate function have the same maximum at $j=0$. The constant $c$ can be determined from the fact that the approximate function should vanish at $j=\frac{-q}{4}+\frac{1}{2}$. From this condition we get $c=(\frac{4}{q-2})^2$. 

Similarly, for the second interval from j=1 to j=$(\frac{q}{4}+\frac{1}{2})$, we approximate the function as $(1-b(j-1)^2) e^{(J r \beta l \cos^2{\frac{\pi}{q}})}$ with $b=(\frac{4}{q-2})^2$.
Thus, we get the following:
\begin{widetext}
\begin{equation}
\int_{-\frac q2+1}^{\frac q2} e^{\beta {Jrl \cos{\frac{\pi}{q}} \cos{\frac{(2\pi j- \pi)}{q}}}} dj \to \int_{\frac{-q}{4}+\frac{1}{2}}^{0} (1-c j^2) e^{(J r \beta l \cos^2{\frac{\pi}{q}})} dj + \int_{1}^{\frac{q}{4}+\frac{1}{2}} (1-b(j-1)^2) e^{(J r \beta l \cos^2{\frac{\pi}{q}})} dj.
\end{equation}
\end{widetext}
After performing the integration, we get the following: 
\begin{equation}
\int_{-\frac q2+1}^{\frac q2} e^{\beta {Jrl\cos{\frac{\pi}{q}}\cos{\frac{(2\pi j- \pi)}{q}}}} dj\to \frac{q-2}{3}e^{(J r \beta l \cos^2{\frac{\pi}{q}})}.
\end{equation}  

Thus, from Eq. \ref{F_eng5}, we get the following expression for the free energy at the local maximum,
\begin{equation}
F_L=\frac{ J r l^2 \cos^2{\frac{\pi}{q}}}{2}-\frac{\ln{(\frac{q-2}{3} e^{(J r \beta l \cos^2{\frac{\pi}{q}})})}}{\beta},
\end{equation}  
with $l=1$, as obtained earlier. Taking $\cos^2{\frac{\pi}{q}}\approx 1-\frac{\pi^2}{q^2}$ for large $q$, we get
\begin{equation}
F_L=-\frac{1}{2} J r(1-\frac{\pi^2}{q^2}) -\frac{1}{\beta}\ln{(\frac{q-2}{3})}.
\label{F_L}
\end{equation} 

From Eqs. \ref{F_G} and \ref{F_L}, we now get an estimate for the energy barrier: 
\begin{equation}
F_L-F_G=\frac{Jr \pi^2}{2 q^2}-\frac{\ln{\frac{q-2}{3}}}{\beta}+\frac{\ln{\frac{q}{3}}}{\beta}.
\label{Bar_eng}
\end{equation}

\subsubsection{Calculation of thermal energy}
To calculate the thermal energy ($U-F$) at a global minimum, we use the following thermodynamic relation to first calculate the internal energy at the global minimum ($U_G$): 
\begin{equation}
    U_G=-T^2 \frac{\partial}{\partial T}(\frac {F_G}{T}),
\end{equation}
where $F_G$ is given by Eq. \ref{F_G}. A simple calculation shows that 
\begin{equation}
    U_G=-\frac{Jr}{2}.
\end{equation}
Hence, the thermal energy at the global minimum is 
\begin{equation}
U_G-F_G=\frac{1}{\beta}\ln (\frac q3).
\label{Th_eng}
\end{equation}

\subsubsection{Estimation of $T_{SSB}$}
As discussed earlier, an estimation of the spontaneous symmetry-breaking transition temperature ($T_{SSB}$) can be found by equating the energy barrier between two consecutive global minima and the thermal energy at a global minimum. Accordingly, from Eqs. \ref{Bar_eng} and \ref{Th_eng}, we get 
\[\frac{Jr \pi^2}{2 q^2}-\frac{\ln{\frac{q-2}{3}}}{\beta}+\frac{\ln{\frac{q}{3}}}{\beta}=\frac{1}{\beta}\ln (\frac q3).\]
Solution of this equation gives (after replacing $\beta$ by $\frac{1}{k_B T_{SSB}}$),
\begin{equation}
T_{SSB}=\frac{Jr\pi^2}{2k_Bq^2 \ln\frac{q-2}{3}}.
\end{equation}
For large-$q$, $q^2$  dominates over $\ln{\frac{q-2}{3}}$, so we see that $T_{SSB}$ decreases as $q^{-2}$, as expected earlier for large-$q$. 
 
 The zeroth order mean-field theory that we discussed provides many insights into the $q$-state clock model. However, this theory does not allow us to estimate the important quantities like spin-spin correlation. The spin-spin correlation helps us gain more insight into the phases of the model. To calculate this quantity, we now develop a higher-order (first order) mean-field theory where two neighboring spins with exact interaction are targeted. This higher-order theory also helps us improve some of the results we already obtained using the zeroth order theory. 
 
\section{Higher-order mean-field theory} \label{sec3}
\begin{figure}
    \centering
    \includegraphics[width=7.5cm, height=4.5cm]{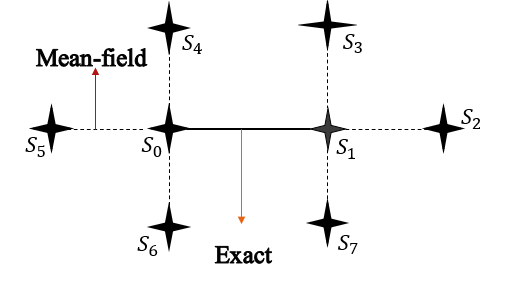}
    \caption{First order mean-field configuration}
    \label{fig:NMF}
\end{figure}
Let us consider a two-dimensional square lattice and focus on the two spins ($S_0$ and $S_1$) that interact exactly. These two spins are in the mean field created by their neighboring spins ($S_2, S_3, S_4, S_5, S_6, S_7 $), as shown in Fig. \ref{fig:NMF}. The Hamiltonian for the system of two spins can be written as: 
\vspace{-1.2mm}
\begin{equation}
\begin{split}
    h_{01}= & -J Re(S_0 S^*_1 \\
            &+\frac{1}{2}(S_5 S^*_0+S_1 S^*_2+S_6 S^*_0 +S_0 S^*_4 +S_7 S^*_1+S_1 S^*_3)).
\end{split}    
\label{h01}
\end{equation}
The factor 1/2 is to avoid double counting the interactions between the targeted spins and their neighbors. If there are $N$ number of spins, then there are $\frac{N}{2}$ such spin pairs. We recall here that the complex spin variable ($S_k$) associated with the $k$-th spin is by definition: $S_k = e^{i\theta_k}$, where $\theta_k=2\pi j_k/q$. The variable $j_k$ takes any integer value from 1 to $q$. To develop the MF theory, we now replace a spin variable with a complex mean-field parameter ($l$) and a small fluctuation or deviation term: $S_k=l+\delta S_k$. We take a translationally invariant system for which the MF parameter $l$ is the same for all the spins. Using the parameter $l$, any interaction term can be approximated as: $S_iS_j^*\approx -|l|^2+l^*S_i+lS_j^*$. With the help of this expression, we get the following mean-field version of Hamiltonian in Eq. \ref{h01},
\begin{equation}
\begin{split}
h_{MF} =& -J Re(S_0S_1^*  \\
        &-\frac 32 |l|^2+\frac 32 l^*S_0+\frac 12 \sum_{j\in N_0} lS_j^*\\
        &-\frac 32 |l|^2+\frac 32 l^*S_1+\frac 12 \sum_{k\in N_1} lS_k^*).
\end{split}
\label{hmf}
\end{equation}
Here, $N_0$ denotes the set of nearest neighbors of the 0-th spin except the 1st spin. Similarly, $N_1$ is the set of nearest neighbors of the 1st spin except the 0-th spin. It may be mentioned here that since we are interested in the real part of an interaction term, terms like $l^*S_0$ and $lS_0^*$ can be taken to be equivalent and can be written interchangeably. From Eq. \ref{hmf}, if we now write the mean-field Hamiltonian for the whole two-dimensional system, we will have exact interaction terms like $S_0S_1^*$ corresponding to each target spin pair and the terms like $l^*S_0$ appearing 3 times in the Hamiltonian. Therefore, the total mean-field Hamiltonian of the system can be written as the sum of $N/2$ mean-field pair Hamiltonians; this mean-field pair Hamiltonian corresponding to $i-j$ pair (nearest neighbors) is given by,
\begin{equation}
h_{ij} = -J Re(S_iS_j^* -3 |l|^2+ 3 l^*S_i+3lS_j^*).
\label{hpair}
\end{equation}
Now we recall that $S_k=\cos{\theta_k} + i \sin{\theta_k}$, where $\theta_k=2\pi j_k/q$ with $j_k$ taking values from 1 to $q$. If we write $l=l_1+i l_2$, we get from Eq. \ref{hpair}:
\begin{equation}
\begin{split}
h_{ij} =& -J \cos{(\theta_i-\theta_j)} + 3J(l_1^2+l_2^2) \\
        &- 3Jl_1(\cos{\theta_i} + \cos{\theta_j})-3Jl_2(\sin{\theta_i}+\sin{\theta_j}).  
\end{split}
\label{hpair1}
\end{equation}
The possible energy values of the spin pair can be obtained from Eq. \ref{hpair1} by taking different possible values of $\theta_i$ and $\theta_j$. We can now calculate the partition function of the system as $Z=Z_{ij}^{\frac N2}$, where $Z_{ij}$ is the partition function corresponding to the pair $i-j$ and is given by,
\begin{equation}
 Z_{ij}=\sum_{\theta_i,\theta_j}e^{-\beta h_{ij}}.   
\end{equation}
Here, the sum is over possible values of $\theta_i$ and $\theta_j$. The free energy corresponding to a spin pair is given by,
\begin{equation}
 F=-\frac{1}{\beta}\ln Z_{ij}.   
\end{equation}
Written explicitly, we get the following expression for the free energy,
\begin{widetext}
\begin{equation} 
F= 3J(l_1^2+l_2^2) -
   \frac{1}{\beta}\ln\sum_{i=1}^{q}\sum_{j=1}^{q} e^{J\beta(\cos{\frac{2 \pi (i-j)}{q}}+3 l_1 (\cos{\frac{2 \pi i}{q}+\cos{\frac{2 \pi j}{q}})+3 l_2(\sin{\frac{2 \pi i}{q}}+\sin{\frac{2 \pi j}{q}})})}.
   \label{F_nmf}
\end{equation}
\end{widetext}

\subsection{Free energy plots at different temperatures}
\begin{figure*}
{\xincludegraphics[scale=0.3,label=\text{(a)}]{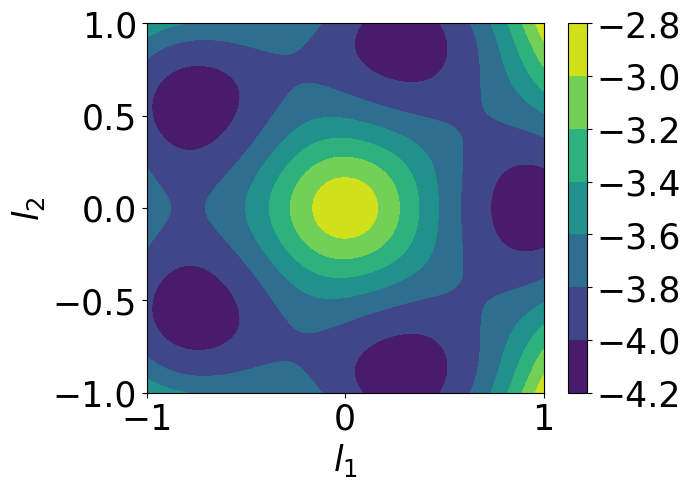}
        \phantomsubcaption\label{2a}}
{\xincludegraphics[scale=0.3,label=\text{(b)}]{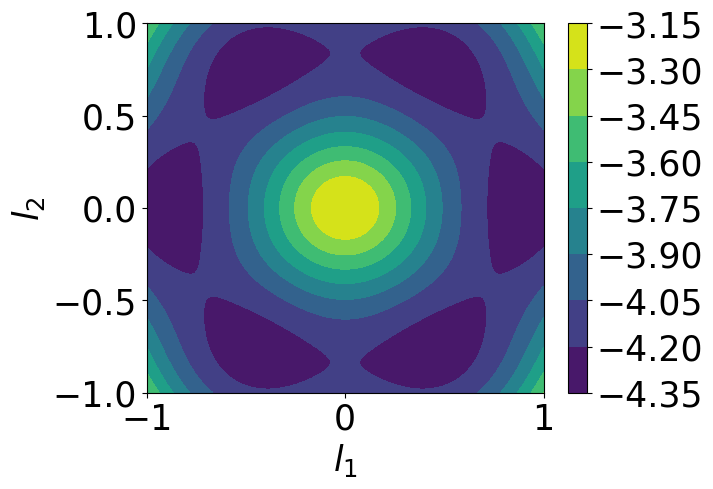}
\phantomsubcaption\label{2b}}
{\xincludegraphics[scale=0.3,label=\text{(c)}]{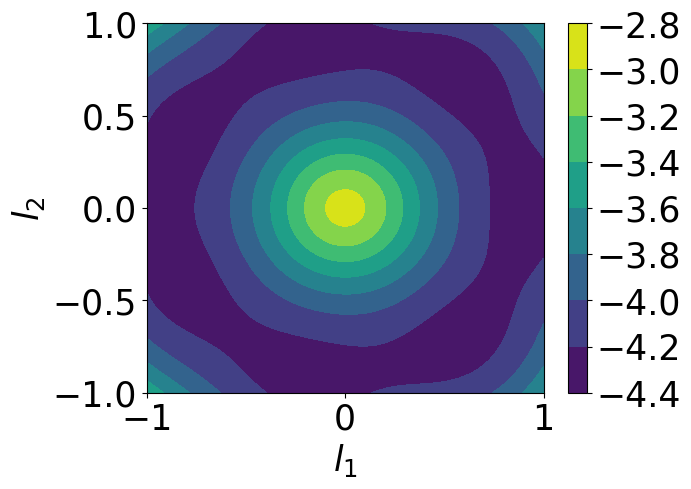}
\phantomsubcaption\label{2c}}

{\xincludegraphics[scale=0.29,label=\text{(d)}]{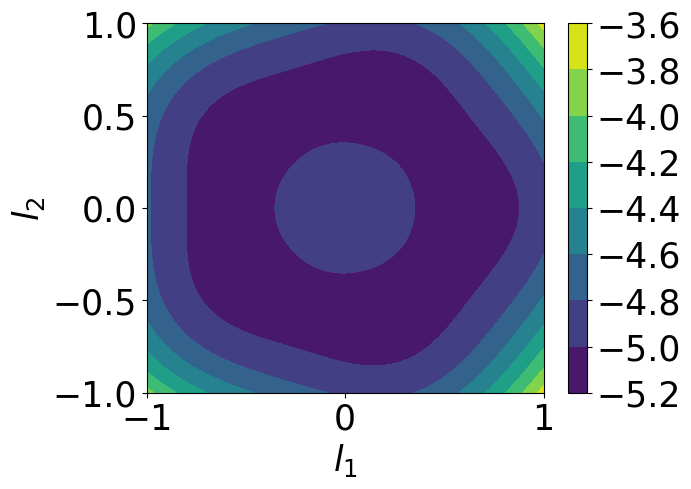}
        \phantomsubcaption\label{2d}}
{\xincludegraphics[scale=0.29,label=\text{(e)}]{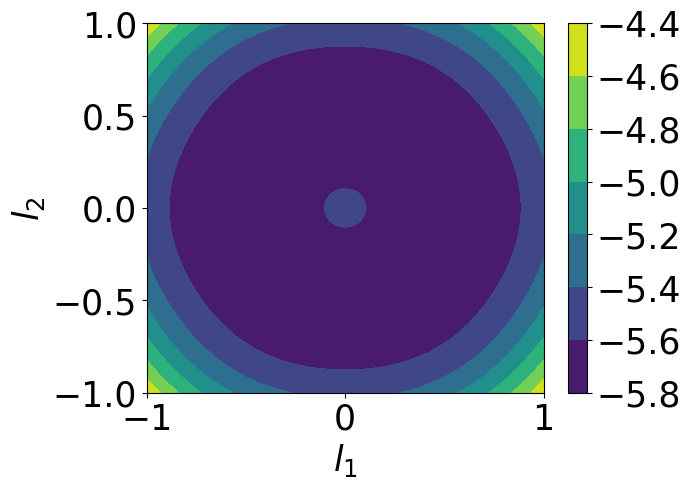}
\phantomsubcaption\label{2e}}
{\xincludegraphics[scale=0.29,label=\text{(f)}]{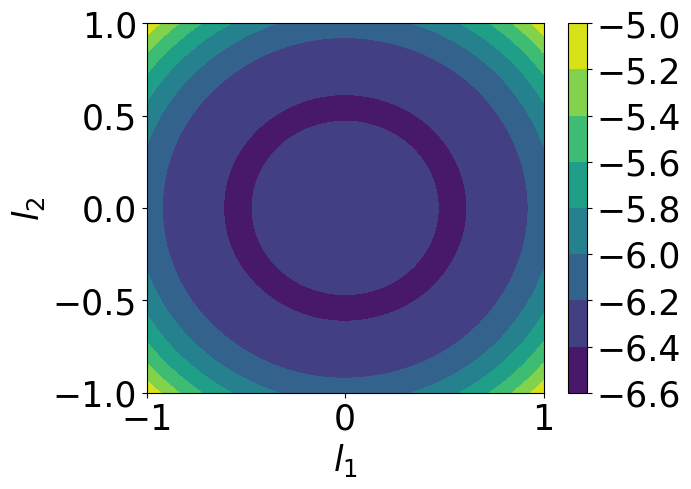}
\phantomsubcaption\label{2f}}

{\xincludegraphics[scale=0.28,label=\text{(g)}]{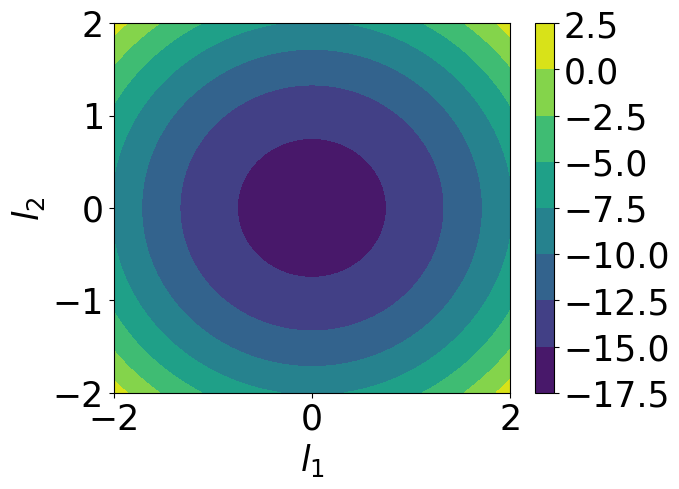}
        \phantomsubcaption\label{2g}}
{\xincludegraphics[scale=0.28,label=\text{(h)}]{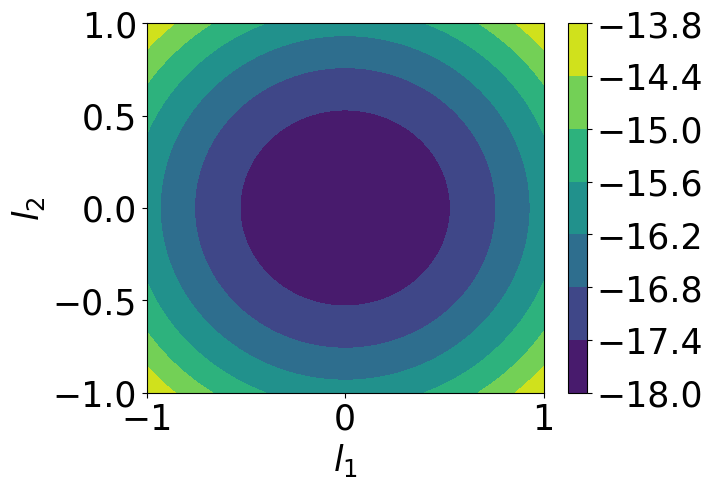}
\phantomsubcaption\label{2h}}
{\xincludegraphics[scale=0.28,label=\text{(i)}]{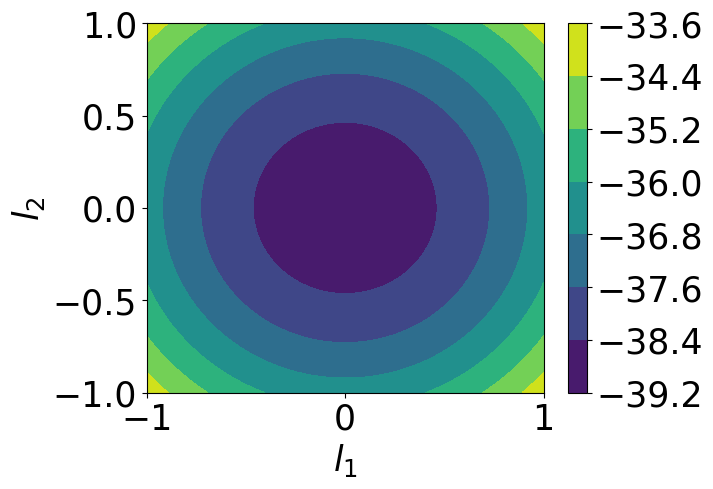}
\phantomsubcaption\label{2i}}
\caption{
First order mean-field results: free energy plots at low, intermediate and high temperatures are shown in (a)-(c), (d)-(f) and (g)-(i), respectively. Results in the left, middle and right columns are for $q=$ 5, 6 and 7, respectively.}
\label{fig:free_plots1}
\end{figure*}
Using this new approximation, the behavior of free energy can be analyzed at a high temperature, intermediate temperature, and low temperature in a similar way as we have done previously (within the zeroth order mean-field approach). The free energy plots in three different temperature ranges for $q=$ 5, 6, and 7 can be found in Fig. \ref{fig:free_plots1}. These new results are qualitatively the same as obtained using the zeroth order theory. There are some quantitative differences, which can be appreciated from the updated value of  $T_{BKT}$, which we again calculate in the following using the new mean-field approach.  

\subsection{Calculation of $T_{BKT}$ using new mean-field approximation} 
We can use the same scheme to calculate the BKT transition temperature as in the first part (Sec. \ref{sec2}). We will find at which temperature the free energy function has an inflection point at (0,0). From Eq. \ref{F_nmf}, we have the free energy as function of $l_1$ ($l_2$=0) as follows:
\begin{equation}
\begin{split}
F(l_1,0) = &~ 3J|l_1|^2  \\    
&-\frac{1}{\beta}\ln\sum_{i=1}^{q}\sum_{j=1}^{q}e^{J\beta(\cos{\frac{2 \pi (i-j)}{q}}+3 l_1 (\cos{\frac{2 \pi i}{q}+\cos{\frac{2 \pi j}{q}}))}}.
\end{split}
\label{F_nmf1}
\end{equation}
Now taking the second order partial derivative of $F(l_1,0)$ at $l_1=0$ and equating it to 0, we obtain:
\begin{equation}
    \begin{split}
        &~ \pdv[order={2}]{F}{l_1}_{l_1=0}= 6J\\
        &+9 \beta J^2\frac{\left(\sum_{i=1}^{q}\sum_{j=1}^{q} (\cos{\frac{2\pi i}{q}}+\cos{\frac{2 \pi j}{q}}) e^{\beta J \cos{\frac{2 \pi (i-j)}{q}}}\right)^2}{\left(\sum_{i=1}^{q}\sum_{j=1}^{q} e^{\beta J\cos{\frac{2 \pi (i-j)}{q}}}\right)^2} \\
        &- 9 \beta J^2\frac{\sum_{i=1}^{q}\sum_{j=1}^{q} (\cos{\frac{2\pi i}{q}}+\cos{\frac{2 \pi j}{q}})^2 e^{\beta J \cos{\frac{2 \pi (i-j)}{q}}}}{\sum_{i=1}^{q}\sum_{j=1}^{q} e^{\beta J\cos{\frac{2 \pi (i-j)}{q}}}}=0.
    \end{split}
\label{eqn_tbkt}
\end{equation}
It is a nontrivial task to find $\beta$ for which the above equation (Eq. \ref{eqn_tbkt}) is satisfied. To go further, we perform the following approximation. As seen in the first part, in Sec. \ref{sec2}, $\beta J \sim \frac 12$ near the BKT transition temperature. Therefore, near the transition temperature, $|\beta J\cos{\frac{2\pi(i-j)}{q}}|\lesssim \frac 12$. Accordingly, we take the following first order approximation: $e^{\beta J \cos{\frac{2\pi(i-j)}{q}}} \approx 1+ \beta J \cos{\frac{2\pi(i-j)}{q}}$. Noting that $\sum_{i=1}^{q} \cos{\frac{2 \pi i}{q}}=0$ and $\sum_{i=1}^{q} \cos^2{\frac{2 \pi i}{q}}=\frac{q}{2}$, we get the following:
\begin{equation}
\begin{split}
&\frac{\sum_{i=1}^{q}\sum_{j=1}^{q} (\cos{\frac{2\pi i}{q}}+\cos{\frac{2 \pi j}{q}})^2 e^{\beta J \cos{\frac{2 \pi (i-j)}{q}}}}{\sum_{i=1}^{q}\sum_{j=1}^{q} e^{\beta J\cos{\frac{2 \pi (i-j)}{q}}}}\\ 
&=\frac{\sum_{i=1}^{q}\sum_{j=1}^{q} (\cos{\frac{2\pi i}{q}}+\cos{\frac{2 \pi j}{q}})^2 (1+\beta J \cos{\frac{2 \pi (i-j)}{q}})}{\sum_{i=1}^{q}\sum_{j=1}^{q} (1+\beta J\cos{\frac{2 \pi (i-j)}{q}})}\\
&=\frac{q .\frac{q}{2}+q .\frac{q}{2} +2 \beta J. \frac{q}{2} . \frac{q}{2}}{q.q}\\
&=1+\frac{\beta J}{2}.
\end{split}
\end{equation}
Using this result, we get from Eq. \ref{eqn_tbkt},
\begin{equation}
\begin{split}
&6J + 9\beta J^2\times 0 - 9\beta J^2 (1+\frac{\beta J}{2}) =0,~\text{i.e.,}\\
&4-6\beta J - 3 (\beta J)^2 = 0.
\label{eqn_tbkt1}
\end{split}
\end{equation}
Taking only the positive root of Eq. \ref{eqn_tbkt1}, we get the following value for the transition temperature (after replacing $\beta$ by $\frac{1}{k_B T_{BKT}}$),
\begin{equation}
T_{BKT}=\frac{1.895J}{k_B}.
\end{equation}

\begin{figure*}
    \centering
{\xincludegraphics[scale=0.25,label=\text{(a)}]{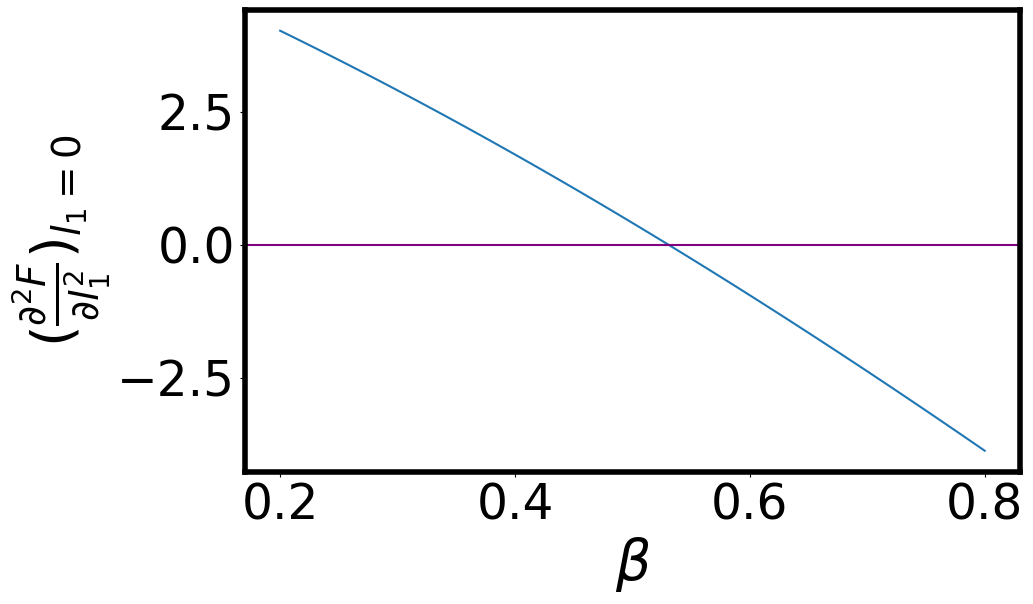}
        \phantomsubcaption\label{a}}
        {\xincludegraphics[scale=0.25,label=\text{(b)}]{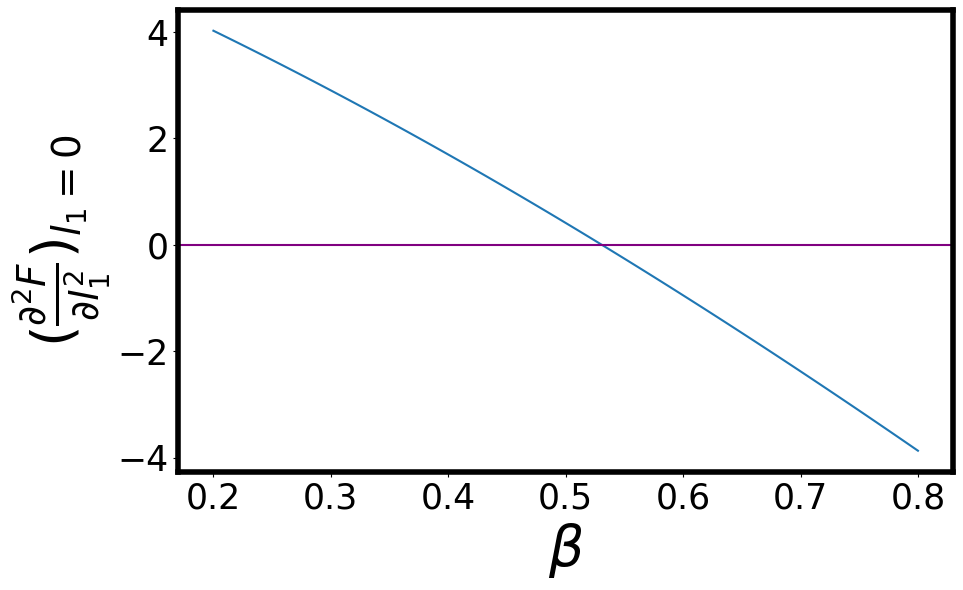}
        \phantomsubcaption\label{b}}
        \caption{Plots of $\pdv[order={2}]{F}{l_1}_{l_1=0}$ vs. $\beta$ 
        for (a) $q=6$ and (b) $q=7$.
   \label{fig:imageBKT}}  
\end{figure*}

Instead of solving Eq. \ref{eqn_tbkt} analytically by taking some approximation, it is also possible to directly solve the equation numerically. For given $q$, we can numerically check at which $\beta$ the equation is satisfied. For $q=$ 6 and 7, plots of $\pdv[order={2}]{F}{l_1}_{l_1=0}$ vs. $\beta$ can be seen Fig. \ref{fig:imageBKT}. Numerical value of $T_{BKT}$ calculated from such plots show that, for $q=7$ or more, the value of the transition temperature does not change much with $q$ and this constant value is estimated to be $T_{BKT}=\frac{1.895J}{k_B}$.
We note that, compared to the zeroth order result, the current value of $T_{BKT}$ is slightly closer to the reported value of $ 0.893 J/ k_B$ as found using Monte Carlo calculations \cite{hsieh13}. Thus, we see that the consideration of a single exact interaction in the higher-order mean-field theory improves the results quantitatively. 
 \begin{figure*}
    \centering
   {\xincludegraphics[scale=0.25,label=\text{(a)}]{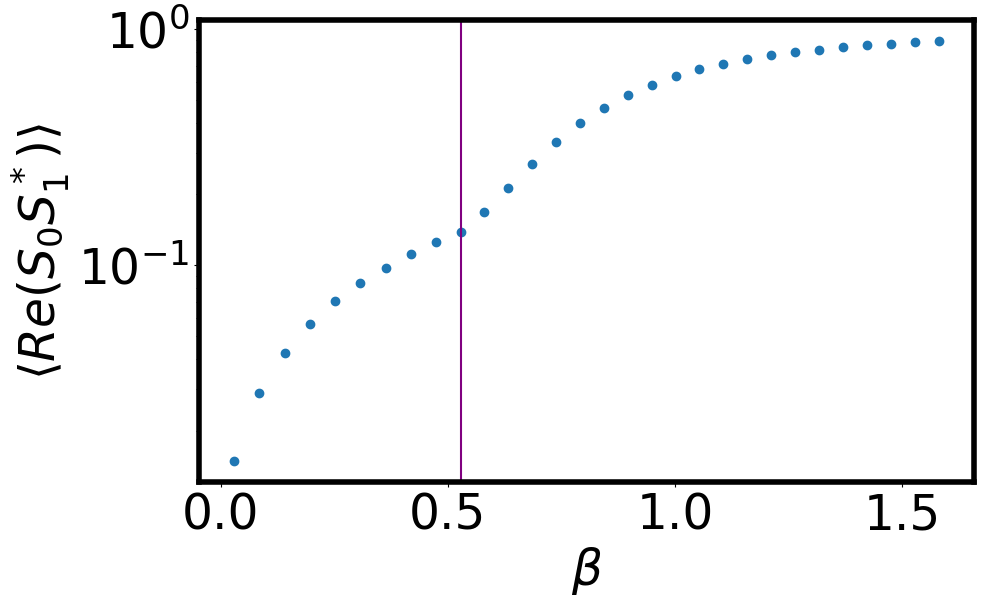}
        \phantomsubcaption\label{fig:correl6}}
        {\xincludegraphics[scale=0.25,label=\text{(b)}]{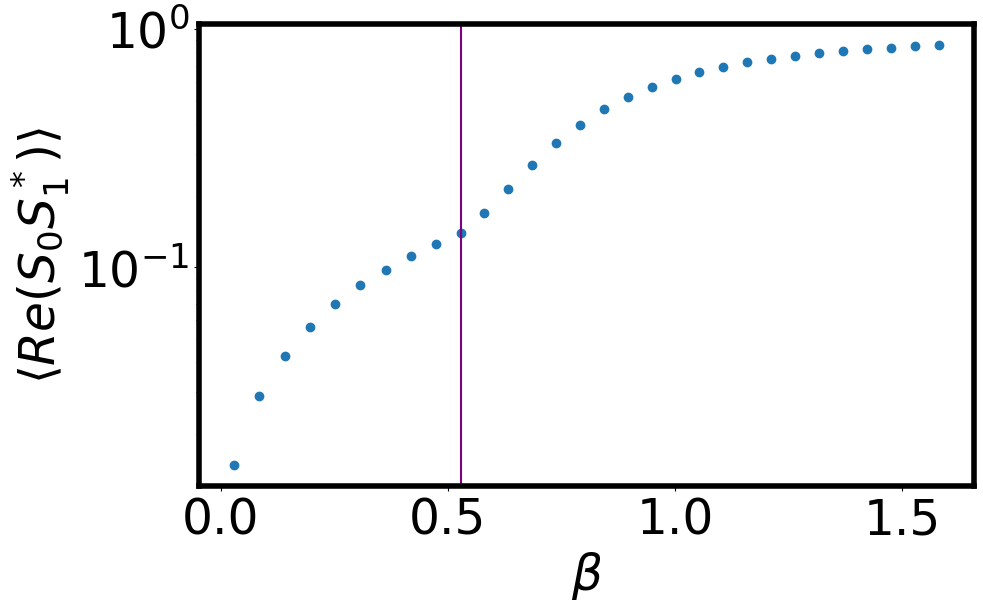}
        \phantomsubcaption\label{fig:correl7}}
    \caption{NN spin-spin correlation for 
    (a) $q=6$, and (b) $q=7$.}
  \label{fig:image2}
  \end{figure*}  

\subsection{Estimation of the spin-spin correlation} \label{nn_corr}
The spin-spin correlation is an important quantity in studying different phases of a system. The correlation shows different behaviors in different phases, and a change in its behavior normally indicates a phase transition. Mean-field theories, in general, do not predict exact correlation behavior. Our first-order mean-field theory considers one exact interaction between two neighboring spin pairs. As a result, it is possible to estimate the nearest-neighbor (NN) spin-spin correlation using this approach. The NN correlation can tell us whether the system is in an ordered or disordered phase at a given temperature. Using our first-order MF theory, we estimate the NN spin-spin correlation for the $q$-state clock model. We note that the NN spin-spin correlation is given by,
\begin{equation}
    \langle Re(S_0 S^*_1) \rangle= \frac{\sum_{\theta_0}\sum_{\theta_1} Re(S_0 S^*_1) e^{-\beta h_{01}}}{\sum_{\theta_0}\sum_{\theta_1} e^{-\beta h_{01}}}.
\label{corr1}
\end{equation}
Here $h_{01}$ is given by Eq. \ref{hpair1} with indices $i$ and $j$ taking values 0 and 1 respectively. In terms of variables $\theta_0$ and $\theta_1$, we also have $Re(S_0 S^*_1)=\cos{(\theta_0-\theta_1)}$. Evaluating this correlation analytically is, in general, difficult. It is, though, possible to numerically evaluate the value of NN correlation at different values of $\beta$. In the following we first discuss how one can calculate NN correlation numerically, we then discuss how one can get analytical results in two different limits (high and low temperature limits).

\subsubsection{Numerical calculation of NN correlation}
To calculate NN correlation as function of $\beta$, we first need to find the appropriate numerical values of the parameters $l_1$ and $l_2$ at different $\beta$ values, since these parameters appear in Eq. \ref{corr1} via $h_{01}$.   We note from the free energy plots (Fig. \ref{fig:free_plots1}) that there is always a minimum on the positive $l_1$-axis (including origin). In the high-temperature limit, the minimum is at the origin ($l_1=l_2=0$). In the low-temperature limit, there are $q$ minima around the origin; one of these minima is on the positive $l_1$-axis. In the intermediate temperature, there is a ring of minima around the origin. This observation allows us to take $l_2=0$ and an appropriate value of $l_1$ (corresponding to the minimum on the positive $l_1$-axis) in $h_{01}$. To find this appropriate value of $l_1$ at a given $\beta$, we numerically find the free energy minimum on the positive $l_1$-axis from the equation $\pdv{F(l_1,0)}{l_1}=0$, where $F(l_1,0)$ is given by Eq. \ref{F_nmf1}. 

For $q=6$ and 7, we calculate $l_1$
corresponding to the minimum on the $l_1$-axis for different $\beta$ values and then use that to calculate NN correlation at those $\beta$ values. Results can be seen in Fig. \ref{fig:image2}. We observe from the plots of NN correlation that there is a kink near $T_{BKT}=\frac{1.895J}{k_B}$. This result correctly predicts a change in phase at $T_{BKT}$. We, though, do not observe any noticeable change in correlation at $T_{SSB}$. 

\subsubsection{Analytical expression of NN correlation in high-temperature limit}
We know from the free energy plots (Fig. \ref{fig:free_plots1}) that, in the high-temperature limit, there is only one minimum corresponding to $l_1=l_2=0$. Accordingly, we can evaluate Eq. \ref{corr1} by taking $l_1=l_2=0$ in the Hamiltonian $h_{01}$. Thus, in this temperature limit, $h_{01}=-J\cos{(\theta_0-\theta_1)}$ and $e^{-\beta h_{01}} \approx 1+\beta J \cos{(\theta_0-\theta_1)}$, since $\beta$ is small. Written explicitly, we get from Eq. \ref{corr1},
\begin{equation}
\begin{split}
    \langle Re(S_0 S^*_1) \rangle=& \frac{\sum_{\theta_0}\sum_{\theta_1} \cos{(\theta_0-\theta_1)}(1+\beta J \cos{(\theta_0-\theta_1)})}{\sum_{\theta_0}\sum_{\theta_1}(1+\beta J \cos{(\theta_0-\theta_1)}) }\\
     =& \frac{\sum_{i=1}^{q}\sum_{j=1}^{q} \cos{\frac{2\pi (i-j)}{q}}(1+\beta J \cos{\frac{2\pi (i-j)}{q}})}{\sum_{i=1}^{q}\sum_{j=1}^{q}(1+\beta J \cos{\frac{2\pi (i-j)}{q}}) }.
\label{corr2}
\end{split}
\end{equation}
Noting the identities $\sum_{i=1}^{q} \cos{\frac{2 \pi i}{q}}=0$ and $\sum_{i=1}^{q} \cos^2{\frac{2 \pi i}{q}}=\frac{q}{2}$, we finally get
\begin{equation}
\langle Re(S_0 S^*_1) \rangle= \frac{\beta J}{2}.
\label{corr3}
\end{equation}
We note here that, in the high-temperature limit, the NN spin-spin correlation decreases with increasing temperature. In accordance with our analytical result, we also note from Fig. \ref{fig:image2} that the numerically calculated NN correlation is linear with $\beta$ near the origin. 

\subsubsection{Analytical expression of NN correlation in low-temperature limit} 
In the low-temperature limit ($\beta $ being large), $l_1 \approx 1$ corresponds to the free energy minimum on the positive $l_1$ axis. In this limit, it is not possible to approximate $e^{-\beta h_{01}}$ by retaining a couple of terms from its (Taylor) series. To go further, for large q, we transform the summation to integration. With $l_1=1$ and $l_2=0$ in Eq. \ref{hpair1}, we have 
\begin{equation}
h_{01}=-J\cos{({\theta_0-\theta_1})}+3J-3J(\cos{\theta_0}+\cos{\theta_1}),
\label{h01_crr}
\end{equation}
with the indices $i$ and $j$ taking their values 0 and 1, respectively. In the large $q$ limit and taking symmetric ranges for the integration variables $\theta_0$ and $\theta_1$, we get from Eq. \ref{corr1}, 
\begin{equation}
    \langle Re(S_0 S^*_1) \rangle= \frac{\int_{\theta_0=-\pi}^{\pi}\int_{\theta_1=-\pi}^{\pi}\cos{(\theta_0-\theta_1)}e^{-\beta h_{01}} d\theta_0 d\theta_1}{\int_{\theta_0=-\pi}^{\pi}\int_{\theta_1=-\pi}^{\pi} e^{-\beta h_{01}}d\theta_0 d\theta_1},
\end{equation}
where $h_{01}$ is given by Eq. \ref{h01_crr}. 
This can be further written as: 
\begin{equation}
    \langle Re(S_0 S^*_1) \rangle= \lim_{x \to 1}\frac{1}{\beta J} \pdv{ }{x}\ln{\left(\int_{-\pi}^{ \pi}\int_{-\pi}^{\pi}E(\theta_0,\theta_1;x)d\theta_0 d\theta_1\right)},
\label{E_eqn}
\end{equation}
where $E(\theta_0,\theta_1;x)=e^{J\beta(x\cos{(\theta_0-\theta_1)}+3(\cos{\theta_0}+\cos{\theta_1}))}$.
We now approximate the function $E(\theta_0,\theta_1;x)$ in a manner similar to how in the first part, we approximated an exponential function by a second-order polynomial. We note that function like $e^{t\cos{\theta}}$, with $t > 0$, has a maximum $e^{t}$ at $\theta=0$. This exponential function effectively vanishes when $\theta\le -\frac{\pi}{2}$ or $\theta\ge \frac{\pi}{2}$. For large $t$, this allows us to approximate the function as $e^{t\cos{\theta}} \approx e^{t(1-\theta^2/2)} \to e^{t}(1-tA\theta^2)$. If $t$ is a given constant, $A$ is determined by the fact that at $\theta =\pm \frac{\pi}{2}$, the approximate function should vanish. If $t$ is a variable, then $A$ is determined by replacing $t$ by its limiting value and making sure that the approximate function vanishes at $\theta= \pm \frac{\pi}{2}$. It may be noted here that the original exponential function and the approximate polynomial function both have the same maximum at $\theta=0$. 

Above discussion helps us to approximate $E(\theta_0,\theta_1;x)$ in the following way:
\begin{equation}
E(\theta_0,\theta_1;x)\to e^{\beta J (x+6)}(1-xA(\theta_0-\theta_1)^2)(1-B\theta_0^2)(1-C\theta_1^2).
\end{equation}
With this approximation, the integration ranges in Eq. \ref{E_eqn} should be taken appropriately. First, we now determine the values of the constants $A$, $B$, and $C$. The constants $B$ and $C$ are determined by making sure that the approximate function should vanish at $\theta_0 = \pm \frac{\pi}{2}$  and $\theta_1 = \pm \frac{\pi}{2}$. This gives us $B=C=(\frac{2}{\pi})^2$. While determining the constant $A$, we note that while $\theta_0=\theta_1$ corresponds to the maximum for both exponential and the approximate polynomial function, both functions should vanish when $|\theta_0-\theta_1|\ge \frac{\pi}{2}$. Since finally we are going to take the limit $x\to 1$, we determine that $A=(\frac{2}{\pi})^2$.  

To find the appropriate ranges of integrations in Eq. \ref{E_eqn} after replacing $E(\theta_0,\theta_1;x)$ by its approximation, we first note that the function $e^{3\beta J \cos{\theta_0}}$ or $e^{3\beta J \cos{\theta_1}}$ effectively vanishes when $|\theta_0|\ge \frac{\pi}{2}$ or $|\theta_1|\ge \frac{\pi}{2}$. On the other hand, the function $e^{x\beta J \cos{(\theta_0-\theta_1)}}$ effectively vanishes when $|\theta_0-\theta_1|\ge \frac{\pi}{2}$. This tells us that, when $-\frac{\pi}{2}\le \theta_0\le 0$, the range of the other variable should be $-\frac{\pi}{2}\le \theta_1\le \frac{\pi}{2} + \theta_0$. Similarly, when $0 \le \theta_0 \le \frac{\pi}{2}$, the range of the other variable should be $-\frac{\pi}{2}+\theta_0\le \theta_1\le \frac{\pi}{2}$.

Let us denote the approximate function by $\tilde{E}(\theta_0,\theta_1;x)$, i.e., 
\begin{equation}
\tilde{E}(\theta_0,\theta_1;x)=e^{\beta J (x+6)}(1-\frac{4x}{\pi^2}(\theta_0-\theta_1)^2)(1-\frac{4\theta_0^2}{\pi^2})(1-\frac{4\theta_1^2}{\pi^2}).
\end{equation}
Now to find the NN correlation using Eq. \ref{E_eqn}, we need to perform the following integration,
\begin{equation}
\begin{split}
I(x)=\int_{\theta_0=-\frac{\pi}{2}}^{0}\int_{\theta_1=-\frac{\pi}{2}}^{\frac{\pi}{2}+\theta_0}\tilde{E}(\theta_0,\theta_1;x)d\theta_0 d\theta_1\\
+\int_{\theta_0=0}^{\frac{\pi}{2}}\int_{\theta_1=-\frac{\pi}{2}+\theta_0}^{\frac{\pi}{2}}\tilde{E}(\theta_0,\theta_1;x)d\theta_0 d\theta_1.
\end{split}
\end{equation}
We find that $I(x)=(3.863-0.9678x)e^{\beta J (x+6)}$. We now have,
\begin{equation}
\begin{split}
\langle Re(S^*_0 S_1) \rangle &= \lim_{x \to 1} \frac{1}{\beta J}\frac{\partial\ln{I(x)}}{\partial x}\\
&=1 - \frac{1}{3\beta J}. 
\end{split}
\end{equation}

We see that the NN spin-spin correlation saturates to the value 1 as the temperature reduces (i.e., $\beta \to \infty$). This analytical result matches well with our numerical results shown in Fig. \ref{fig:image2}. 

\subsubsection{Characterizing phases using NN correlation}
We noticed from our numerical as well as analytical results that $\lim _{\beta \to 0}\langle Re(S^*_0 S_1) \rangle = 0$. This result signifies that, at high-temperature limit, the neighboring spins are not correlated or they move randomly. This high-temperature phase is, then, a disordered phase, as expected.

Similarly, in the other limit, we find that $\lim _{\beta \to \infty}\langle Re(S^*_0 S_1) \rangle = 1$. This result says that, in low-temperature limit, the neighboring spins are maximally correlated. For a ferromagnetic model, this implies a low-temperature ordered phase (all spins orient along the same direction). 

In general, in the intermediate temperature regime, $0<\langle Re(S^*_0 S_1) \rangle <1$. This result says that the neighboring spins are not parallel and they, on average, make some angle with each other. This finding, along with the free energy results as shown in Figs. \ref{fig:free_plots1} (d)-(f), implies that, in the intermediate temperature limit, an individual spin orient along a particular direction but different neighboring spins orient in such a way that we get the vortex-antivortex topological defects (as seen in $XY$ model). Here we also refer to our discussion in Sec. \ref{free_plots_0}. 

\section{Conclusion} \label{sec4}
The $q$-state clock model garnered a lot of interest due to the fact that it shows a double phase transitions for finite $q \ge 5$ in two dimension. There are varying opinions about the nature of phases and associated phase transitions, especially concerning the nature of the transition at lower temperature.  Except for a few special values of $q$, the 2D model is not exactly solvable. Different numerical techniques are developed to gain insights into the properties of this model. To investigate some of the unanswered questions and to address some of the debates in this field, we develop a mean-field theory (both basic and higher order) to study the $q$-state clock model. To the best of our knowledge, this is the first systematic mean-field study of the model with finite $q$. 

Our work provides a comprehensive analysis of the $q$-state clock model in two-dimension. Our findings reaffirm the existence of three different phases: (broken) $\mathbb{Z}_q$ symmetric ferromagnetic (ordered) phase at the low-temperature, emergent $U(1)$ symmetric BKT phase at the intermediate temperature and paramagnetic (disordered) phase at the high-temperature. We find that the transition at the higher temperature is of BKT type. We also argue here that the transition at the lower temperature is of the large-order SSB type. Our MF theory helps us confirm that the transition temperature $T_{BKT}$ does not depend on $q$ value (for large $q$) and the transition temperature $T_{SSB}$ decreases quadratically with $q$.

The higher-order mean-field approach developed here refines our understanding by incorporating exact treatments of nearest-neighbor interactions. This approach enhances our ability to describe the model's behavior with better accuracy (resulting in, for example, slightly better estimation for $T_{BKT}$). Using this approach, we could estimate NN spin-spin correlation, which then helped us better characterize the phases.  
 
Our mean-field method can serve as a robust framework for exploring the complex phase structure of the $q$-state clock model. Future research could build upon the present findings to further investigate the model. For example, our mean-field approach can be used in advancing our understanding of the clock model in higher dimensions or can be used in studying the model under applied field. In fact, there has been a lot of interest in understanding whether there exists any intermediate phase in the three (or higher) dimensional clock model. It was conjectured that the model has a (broken) continuous symmetry in the intermediate temperature region \cite{vanEnter11}. Many numerical results for the model and its variants partially support the conjecture \cite{Miyashita97,Scholten93,Hove03,Ueno93,Shao20,Lou07,Todoroki02}. Our mean-field theory, which is expected to perform better in higher dimensions, suggests that there is an intermediate phase with (broken) $U(1)$ symmetry in three or higher dimensions.

\begin{acknowledgments}
We acknowledge Smitarani Mishra for her participation in discussions in the initial phase of this project. 
\end{acknowledgments}
\vspace{3mm} 

\noindent {\bf Author contributions:} RK and MG calculated the BKT transition temperature using the zeroth order MF theory, AG performed most of the remaining calculations using the zeroth and higher order MF theories, SS proposed and closely supervised the project besides completing some calculations, SS wrote the manuscript with the help from AG. 

\nocite{*}

\bibliography{manuscript}

\end{document}